\documentclass[12pt]{iopart}
\pdfoutput=1
\usepackage{graphicx}
\usepackage[position=top]{subfig}
\usepackage[usenames]{color}
\usepackage{amsfonts}
\usepackage{amssymb}
\usepackage[normalem]{ulem}

\newcommand{\sub}[2]{ _{\mathrm{#1}#2}}
\newcommand{\sect}[1]{\section{#1}}
\newcommand{\ssect}[1]{\subsection{#1}}
\newcommand{\LJ}[1]{ \mathcal{L}_{#1} }
\newcommand{\bm}[1]{ {\mathbf{#1} } }
\newcommand{\ecp}{\varepsilon_\mathrm{cp}}
\newcommand{\ecc}{\varepsilon_\mathrm{cc}}
\newcommand{\scp}{s_\mathrm{cp}}
\newcommand{\scc}{s_\mathrm{cc}}
\newcommand{\gcp}{g_\mathrm{cp}}
\newcommand{\gcc}{g_\mathrm{cc}}
\newcommand{\eb}{\ecc}
\newcommand{\epc}{\ecp}
\newcommand{\kbt}{k_\mathrm{b} T}
\newcommand{\kb}{k_\mathrm{b}}
\definecolor{Blue}{rgb}{0,0.0,1.0}
\definecolor{Red}{rgb}{1.0,0.0,0.0}
\newcommand{\Np}{N_\mathrm{p}}
\newcommand{\lp}{N_\mathrm{p}}
\newcommand{\conc}{c_0}
\newcommand{\logc}{\log\conc}
\newcommand{\tobs}{t_\mathrm{obs}}
\newcommand{\kt}{k_\mathrm{B}T}
\newcommand{\Lp}{\lp}
\newcommand{\Lpmax}{\lp^\mathrm{max}}
\newcommand{\Rg}{R_\mathrm{g}}
\newcommand{\csurf}{c_\mathrm{1}^\mathrm{eq}}
\newcommand{\nanlop}{\mathcal{O}_\mathrm{nanl}}
\newcommand{\Nanl}{\nanlop}
\newcommand{\Nc}{N_\mathrm{C}}
\newcommand{\tnuc}{\tau_\mathrm{nuc}}
\newcommand{\nnuc}{n_\mathrm{nuc}}
\newcommand{\rsub}[1]{_\mathrm{#1}}
\newcommand{\Lpeq}{N_\mathrm{p,eq}}
\newcommand{\change}[1]{{#1}}
\newcommand{\mfhch}[1]{{#1}}
\newcommand{\tnucempty}{\tnuc^\mathrm{empty}}
\newcommand{\css}{c_\mathrm{ss}}

\newcommand{\Kd}{K_\mathrm{d}}

\begin{document}

\title{Encapsulation of a polymer by an icosahedral virus}
\author{Oren M. Elrad and Michael F. Hagan}
\address{Dept. of Physics, Brandeis Unversity, Waltham, MA}
\ead{hagan@brandeis.edu}

\begin{abstract}
The coat proteins of many viruses spontaneously form icosahedral capsids around nucleic acids or other polymers. Elucidating the role of the packaged polymer in capsid formation could promote biomedical efforts to block viral replication and enable use of capsids in nanomaterials applications. To this end, we perform Brownian dynamics on a coarse-grained model that describes the dynamics of icosahedral capsid assembly around a flexible polymer. We identify several mechanisms by which the polymer plays an active role in its encapsulation, including cooperative polymer-protein motions. These mechanisms are related to experimentally controllable parameters such as polymer length, protein concentration, and solution conditions. Furthermore, the simulations demonstrate that assembly mechanisms are correlated to encapsulation efficiency, and we present a phase diagram that predicts assembly outcomes as a function of experimental parameters. We anticipate that our simulation results will provide a framework for designing in vitro assembly experiments on single-stranded RNA virus capsids.
\end{abstract}
\pacs{87.10.Tf,87.15.ap}
\submitto{\PB}
\noindent{\it Keywords\/}: Virus Capsid Self-Assembly Computation Simulation

\maketitle

\section{Introduction}
\label{sec:introduction}
During the replication of many viruses with single-stranded RNA (ssRNA) genomes, hundreds to thousands of protein subunits spontaneously assemble around the viral nucleic acid to form an icosahedral protein shell, or capsid. Understanding the factors that confer robustness to this cooperative multicomponent assembly process would advance technologies that exploit capsids as drug delivery vehicles or imaging agents \cite{Gupta2005,Garcea2004,Dietz2004,Soto2006,Sapsford2006,Boldogkoi2004,Dragnea2003}, and could establish principles for the design of synthetic containers with controllable assembly or disassembly. Furthermore, numerous human pathogenic viruses have ssRNA genomes, and understanding how nucleic acid properties promote capsid assembly could spur the development of antiviral drugs that block viral replication.
The nucleic acid cargo is essential for assembly, since ssRNA viral proteins require RNA (or other polyanions \cite{Hiebert1968,Bancroft1969,Dixit2006,Loo2006,Goicochea2007,Huang2007,Loo2007,Sikkema2007,Sun2007,Hu2008,Comellas-Aragones2009,Crist2009,Chang2008}) to assemble at physiological conditions. However, the role of the packaged polymer is poorly understood because assembly intermediates are transient and thus challenging to characterize with experiments. Therefore, this article considers dynamical simulations of a model for icosahedral capsid assembly around a flexible polymer, which result in experimentally testable predictions for the morphologies and yields of assembly products as functions of polymer length and solution conditions. Furthermore, the simulations demonstrate that, depending on solution conditions and the strength of interactions between viral proteins, assembly around a polymer can proceed by significantly different mechanisms. How the interactions among viral components control their assembly mechanisms and products is a fundamental question of physical virology.

Performing atomistic simulations of the complete dynamics of a capsid assembling around its genome is not computationally feasible \cite{Freddolino2006}. However, experimental model systems in which capsid proteins assemble into icosahedral capsids around synthetic polyelectrolytes \cite{Hiebert1968,Bancroft1969,Sikkema2007,Hu2008,Comellas-Aragones2009}, charge-functionalized nanoparticles  \cite{Dixit2006,Loo2006,Goicochea2007,Huang2007,Loo2007,Sun2007}, and nano-emulsions \cite{Chang2008}  demonstrate that properties specific to nucleic acids are not required for capsid formation or cargo packaging. Therefore, in this article we strive for general conclusions about the assembly of an icosahedral shell around a polymer by considering a simplified geometric model, inspired by previous simulations of empty capsid assembly \cite{Rapaport2004, Nguyen2007}. The model employs trimeric protein subunits, represented as rigid triangular bodies, with short ranged attractions arranged so that an icosahedron is the lowest energy state. The subunits experience short range attractive interactions (representing the effect of screened electrostatics) with a flexible polymer, and assembly is simulated with Brownian dynamics.

By taking advantage of their high degrees of symmetry and structural regularity, the structures of virus capsids assembled around single-stranded nucleic acids have been revealed by x-ray crystallography and/or cryo-electron microscopy (cryo-EM) images  (e.g.\cite{Fox1994, Valegard1997, Johnson2004,Tihova2004,Krol1999,Stockley2007,Toropova2008,Larson2005,Lucas2002,Valegard1994,Worm1998,Grahn2001,Valegard1997,Helgstrand2002,Schneemann2006}). The packaged nucleic acids are less ordered than their protein containers and hence have been more difficult to characterize. However cryo-EM experiments have identified that the nucleotide densities are nonuniform, with a peak near the inner capsid surface and relatively low densities in the interior\cite{Tihova2004,Zlotnick1997,Conway1999}. For some viruses striking image reconstructions show that the packaged RNA adopts the symmetry of its protein capsid (e.g. \cite{Tihova2004,Toropova2008,Schneemann2006}). While atomistic detail has not been possible in these experiments, all-atom models have been derived from equilibrium simulations \cite{Freddolino2006,Devkota2009}. Furthermore, a number of equilibrium calculations have analyzed the electrostatics of packaging a polyelectrolyte inside a capsid \cite{Siber2008,Hu2008c,Devkota2009,Forrey2009,Harvey2009,Belyi2006,Angelescu2006,Schoot2005,Zhang2004a,Lee2008,Angelescu2008,Jiang2009}.

Despite these structural studies and equilibrium calculations, the kinetic pathways by which capsid proteins assemble around their genome or other cargoes remain incompletely understood. An in vitro experiment on assembly of cowpea chlorotic mottle virus (CCMV) \cite{Johnson2004} demonstrated different kinetics than for assembly of capsid proteins alone. The results suggested protein-RNA complexes as important intermediates and showed that the relative concentrations of protein and RNA affect assembly mechanisms. However, the structures of intermediates and the specific assembly mechanisms could not be resolved. Recently several groups have begun to overcome this limitation by characterizing assembly intermediates using mass spectrometry (e.g. \cite{Lorenzen2008,Uetrecht2008,Stockley2007,Toropova2008,Basnak2010}). Stockley and coworkers \cite{Stockley2007,Toropova2008,Basnak2010} performed a remarkable series of experiments on MS2 that, along with a computational study \cite{Dykeman2010}, provide strong evidence that RNA binding allosterically mediates conformational changes that dictate capsid morphologies. However, many assembly intermediates and thus the complete assembly pathways could not be resolved. Furthermore, while experiments have examined the relationship between solution conditions and assembly morphologies for CCMV \cite{Bancroft1970,Adolph1974,Lavelle2009}, the effect of the properties of the nucleic acid cargo, such as its length and interactions with the capsid proteins, on capsid assembly morphologies has received only limited exploration (e.g. \cite{Bancroft1969,Loo2007,Sun2007,Hu2008,Krol1999}).

Theoretical or computational modeling therefore can play an important role in understanding the dynamics of capsid assembly around a polymer and the relationship between polymer properties and the structures that emerge from assembly. Several previous modeling efforts have postulated roles of the RNA in the formation of icosahedral geometries \cite{Stockley2007,Rudnick2005} and in enhancing assembly rates \cite{Hu2007b}, but the final structure and assembly pathways were pre-assumed. Recently our group \cite{Kivenson2010} explored capsid assembly around a flexible polymer with a model defined on a cubic lattice, which allowed simulation of large capsid-like cuboidal shells over long time scales. By simulating assembly with a wide range of capsid sizes and polymer lengths, we found that there is an optimal polymer length which maximizes encapsulation yields at finite observation times. The optimal length scales with the number of attractive sites on the capsid, unless there are attractions between polymer segments.

In this article, we perform dynamical simulations on the encapsulation of a flexible polymer by a model capsid with icosahedral symmetry, which enables the predicted assembly products to be directly compared to experimentally observed morphologies. Depending on polymer length and solution conditions, the simulations predict assembly morphologies that include the polymer completely encapsulated by the icosahedral capsid or non-icosahedral capsules, and several forms of disordered assemblages that fail to completely enclose the polymer. Furthermore, we are able to determine the importance of cooperative subunit-polymer motions, which were poorly supported by the single particle Monte Carlo moves used in \cite{Kivenson2010}.

We find that the relationships between polymer length, interaction strengths, and assembly yields are qualitatively similar to Ref.~\cite{Kivenson2010}, but that a different assembly mechanism emerges when the interactions between capsid subunits are very weak and interactions with the polymer are relatively strong. In this mechanism, first hypothesized by McPherson \cite{McPherson2005} and later by Refs. \cite{Hagan2008,Harvey2009}, a large number of subunits bind to the polymer in a disordered fashion, and then collectively reorient to form an ordered shell. This mechanism can lead to a high yield of well-formed capsids assembled around polymers for carefully tuned parameters, but complete polymer encapsulation is sensitive to changes in system parameters.  Regions of parameter space that support the sequential assembly mechanism known for empty capsid assembly \cite{Zlotnick2005} are more robust to variations in parameters.  Finally, we demonstrate that assembly yields are controlled by a competition between kinetics and thermodynamics by comparing the predictions of our dynamical simulations at finite observation times to the equilibrium thermodynamics for the same model. We find that the thermodynamically optimal polymer length is larger than the optimum found in the dynamical simulations, but that thermodynamics can identify the maximum polymer length at which significant yields are achieved in a dynamics. Understanding the relationship between kinetics and equilibrium predictions could be especially useful because it is possible to perform equilibrium calculations on models with more detail than is feasible with dynamical simulations (e.g. \cite{Schwartz1998,Hagan2006,Hicks2006,Nguyen2007,Wilber2007,Nguyen2009,Johnston2010,Wilber2009,Zlotnick1994,Endres2002,Zlotnick2005,Endres2005,Rapaport1999,Rapaport2004,Rapaport2008}).

\change{ Finally, we note that the simulations in this work are meant to represent experimental model systems in which capsid proteins assemble around synthetic polyelectrolytes \cite{Bancroft1969,Sikkema2007,Hu2008} or homopolymeric RNA. This choice was made because: (1) Capsids assemble around synthetic polyelectrolytes \cite{Bancroft1969,Sikkema2007,Hu2008} and nanoparticles \cite{Chen2006,Dixit2006,Sun2007,Goicochea2007,Huang2007}, which demonstrates that properties specific to nucleic acids are not required for capsid formation or cargo packaging. (2) The tertiary structures of viral RNAs in solution are poorly understood \cite{Yoffe2008}. Given the dearth of knowledge about viral RNA base pairing, we consider a simple polymer model that emphasizes universal aspects of capsid assembly around flexible polymers. However, nucleic acid base pairing and sequence dependent interactions could have important effects on assembly pathways and kinetics of assembly around single-stranded RNA; some of these potential effects are highlighted in the context of our simulation results.
}

\sect{Methods}

\ssect{Subunit model}
\label{sec:model}

Capsid proteins typically have several hundred amino acids and assemble on time scales of seconds to hours.  Thus, simulating the spontaneous assembly of even the smallest icosahedral capsid with 60 proteins is infeasible at atomic resolution \cite{Freddolino2006}. However, it has been shown that the capsid proteins of many viruses adopt folds with similar excluded volume shapes, often represented as trapezoids \cite{Mannige2008}. We thus follow the approach taken in recent simulations of the assembly of empty icosahedral shells \cite{Schwartz1998,Hagan2006,Hicks2006,Nguyen2007,Wilber2007,Nguyen2009,Johnston2010,Wilber2009,Rapaport1999,Rapaport2004,Rapaport2008} in which we imagine integrating over degrees of freedom that fluctuate on time scales much shorter than subunit collision times to arrive at simple model for capsid subunits in which they have an excluded volume geometry and orientation-dependent attractions designed such that the lowest energy structure is an icosahedral shell.

\begin{figure}
\begin{center}
\label{fig:model}
\subfloat[]{
    \includegraphics[width=0.425\textwidth]{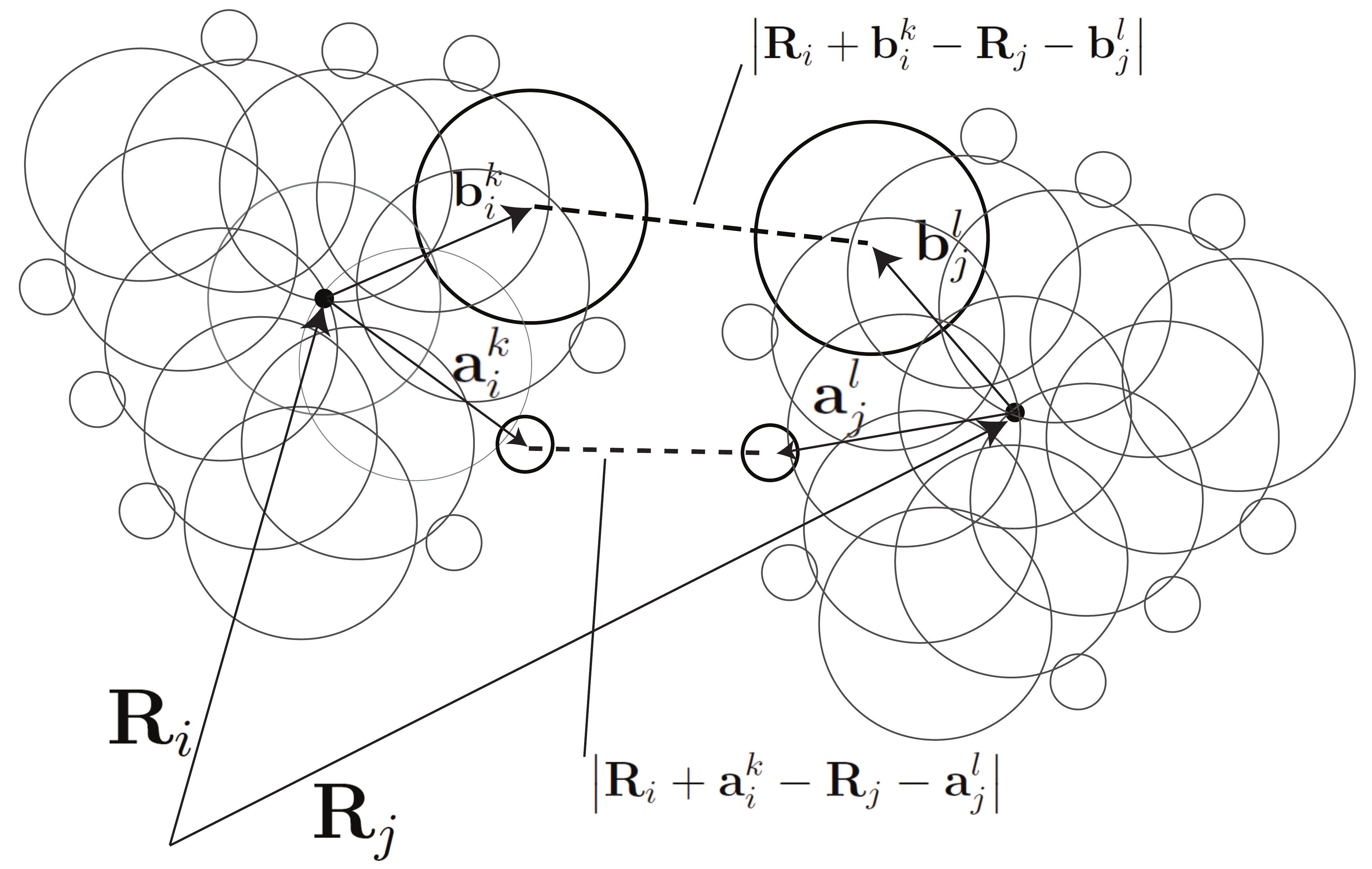}
    \label{fig:model2d}
}
\subfloat[]{
    \includegraphics[width=0.275\textwidth]{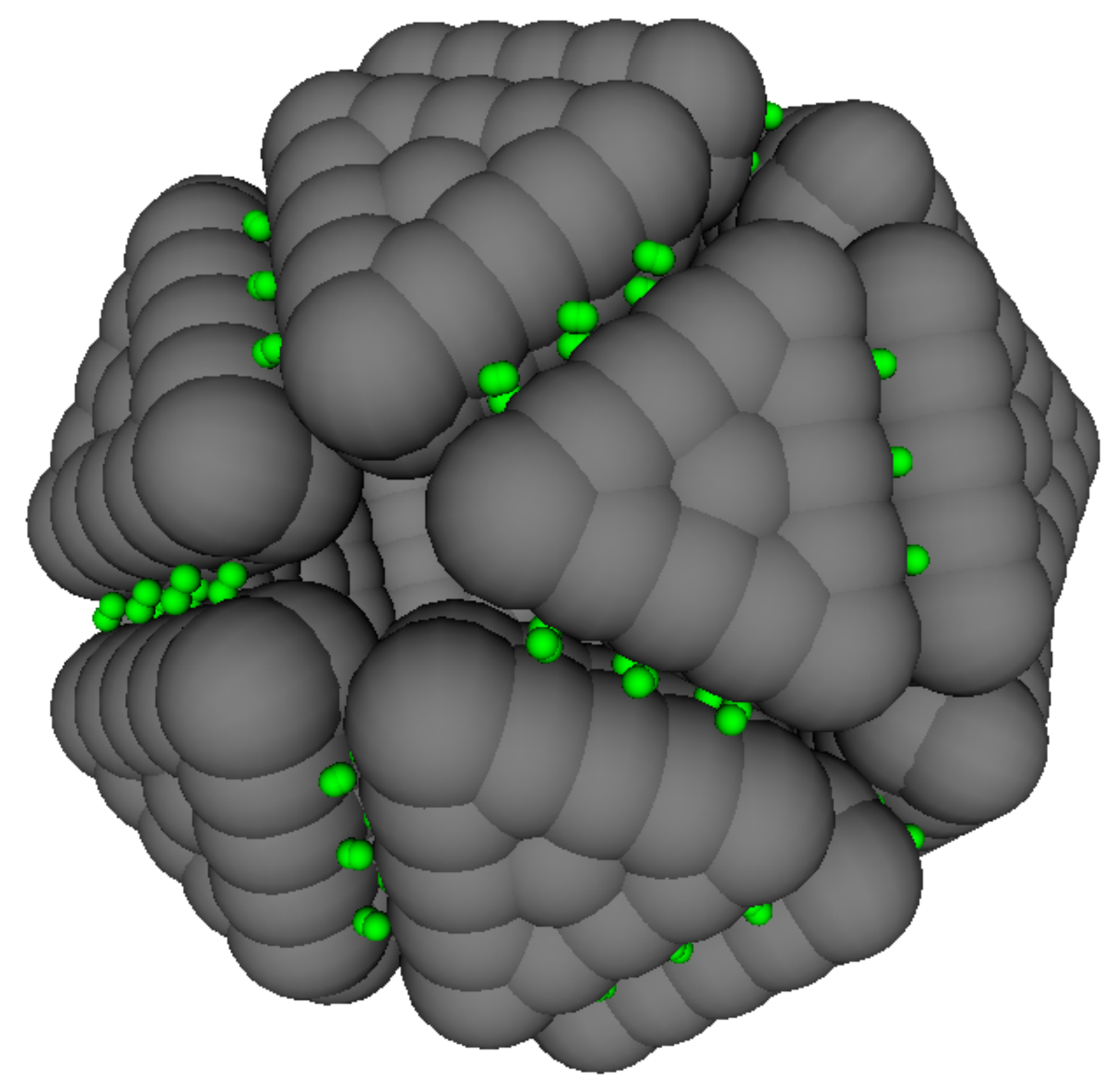}
    \label{fig:modelcapsid}
}
\subfloat[]{
    \includegraphics[width=0.225\textwidth]{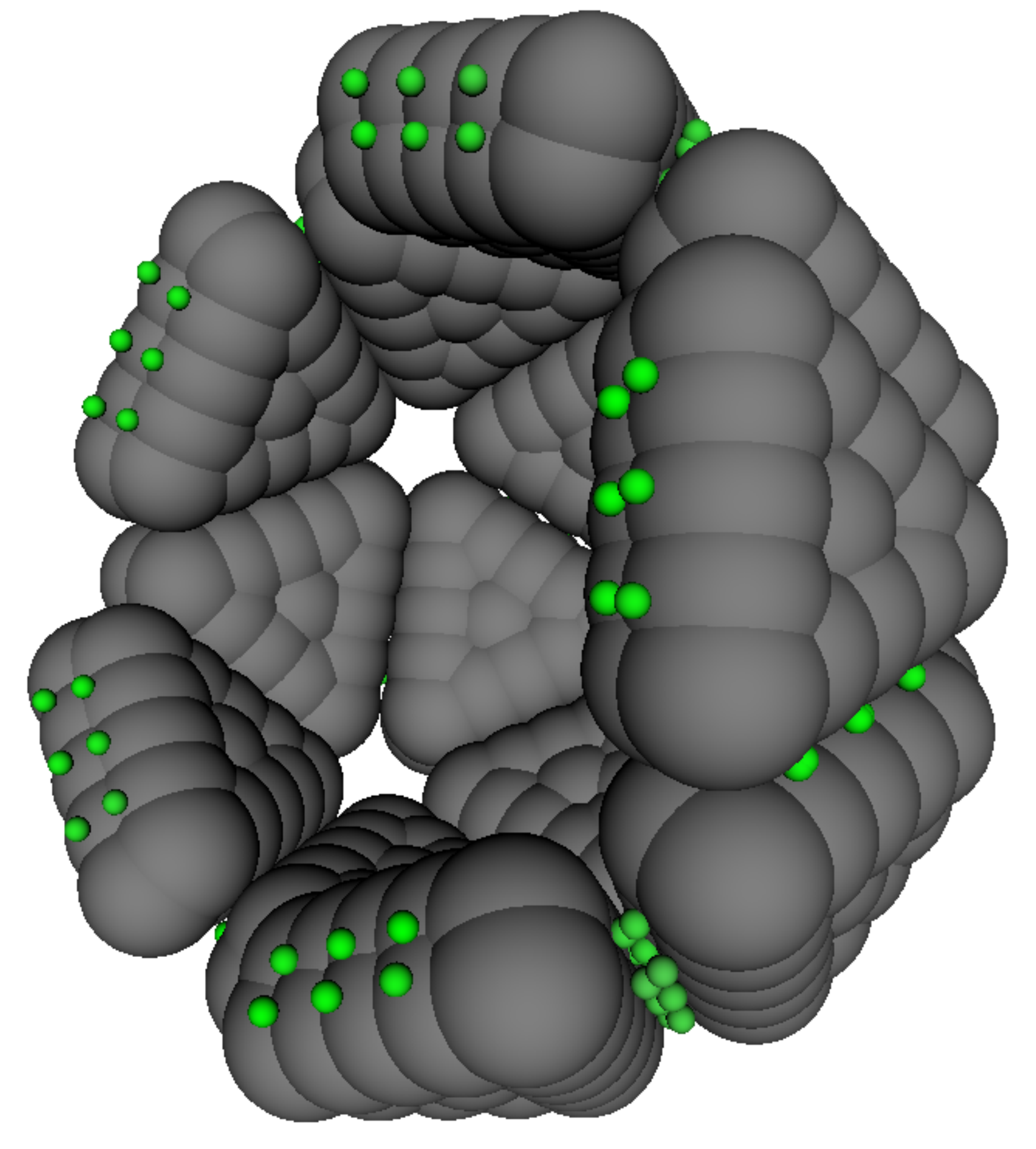}
    \label{fig:modelcapsidcutz}
}
\caption{ The model capsid geometry. {\bf (a) } Two dimensional projection of one layer of a model subunit illustrating the geometry of the capsomer-capsomer pair potential, equation (\ref{Ucc}), with a particular excluder and attractor highlighted from each subunit. The potential is the sum over all excluder-excluder and complementary attractor-attractor pairs. { \bf (b) } An example of a well-formed model capsid. {\bf (c)} Cutaway of a well-formed capsid.
}
\label{fig:modelCapsid}
\end{center}
\end{figure}

Specifically, we consider truncated-pyramidal capsomers designed such that the lowest energy structure is a perfect icosahedron (figure \ref{fig:modelcapsid}). This design is similar to models used by Rapaport et al.\cite{Rapaport1999,Rapaport2004,Rapaport2008} and Nguyen et al.\cite{Nguyen2007} in simulations of empty capsid assembly and could correspond to capsomers comprised of a trimer of proteins that form a T=1 capsid. The model subunits are comprised of a set of overlapping spherical `excluders' that enforce excluded volume and spherical `attractors' with short-range pairwise, complementary attractions that decorate the binding interfaces of the subunit. Each subunit is comprised of two layers of excluders and attractors. Attractor positions are arranged so that complementary attractors along a subunit-subunit interface perfectly overlap in the ground state configuration; excluders on either side of the interface are separated by exactly the cut off of their potential ($x\rsub{c}$, Eq.~\ref{eq:LJ}).  Subunits have no internal degrees of freedom -- they translate and rotate as rigid bodies.

\begin{figure}
\begin{center}
\label{fig:model2}
\subfloat[]{
    \includegraphics[width=0.3\textwidth]{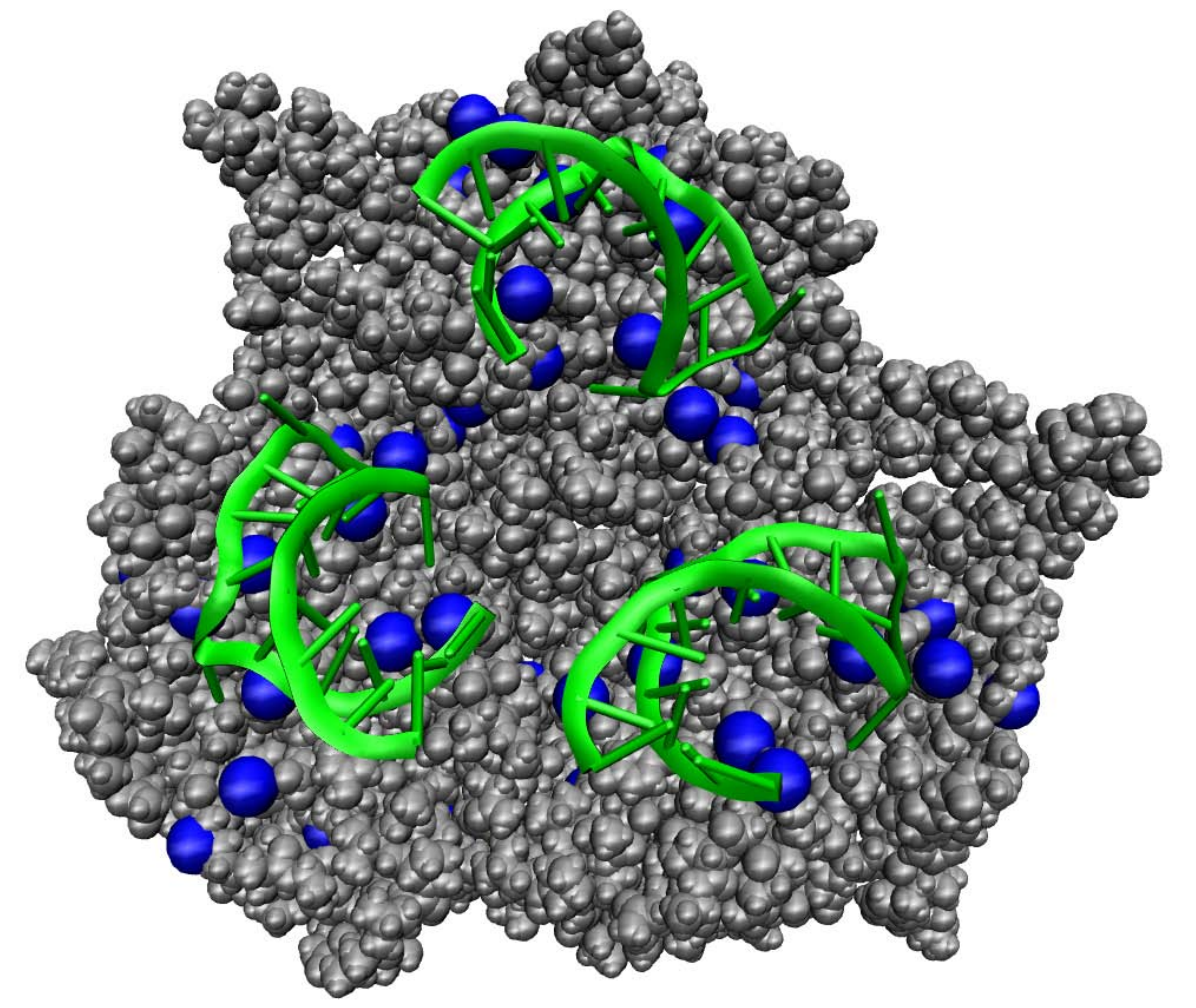}
    \label{fig:MS2}
}
\subfloat[]{
    \includegraphics[width=0.275\textwidth]{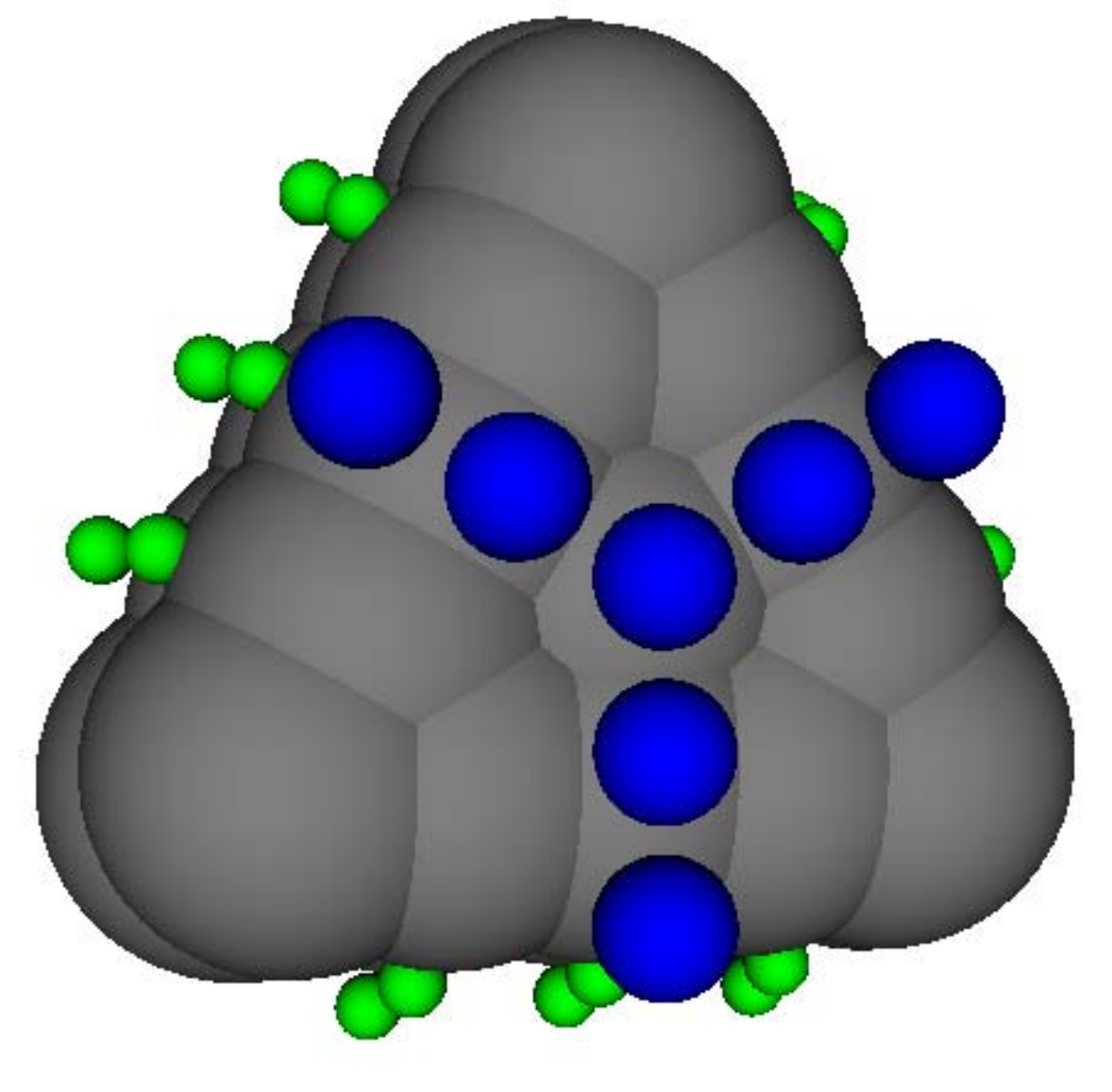}
    \label{fig:modelpolyatt}
}
\subfloat[]{
    \includegraphics[width=0.2625\textwidth]{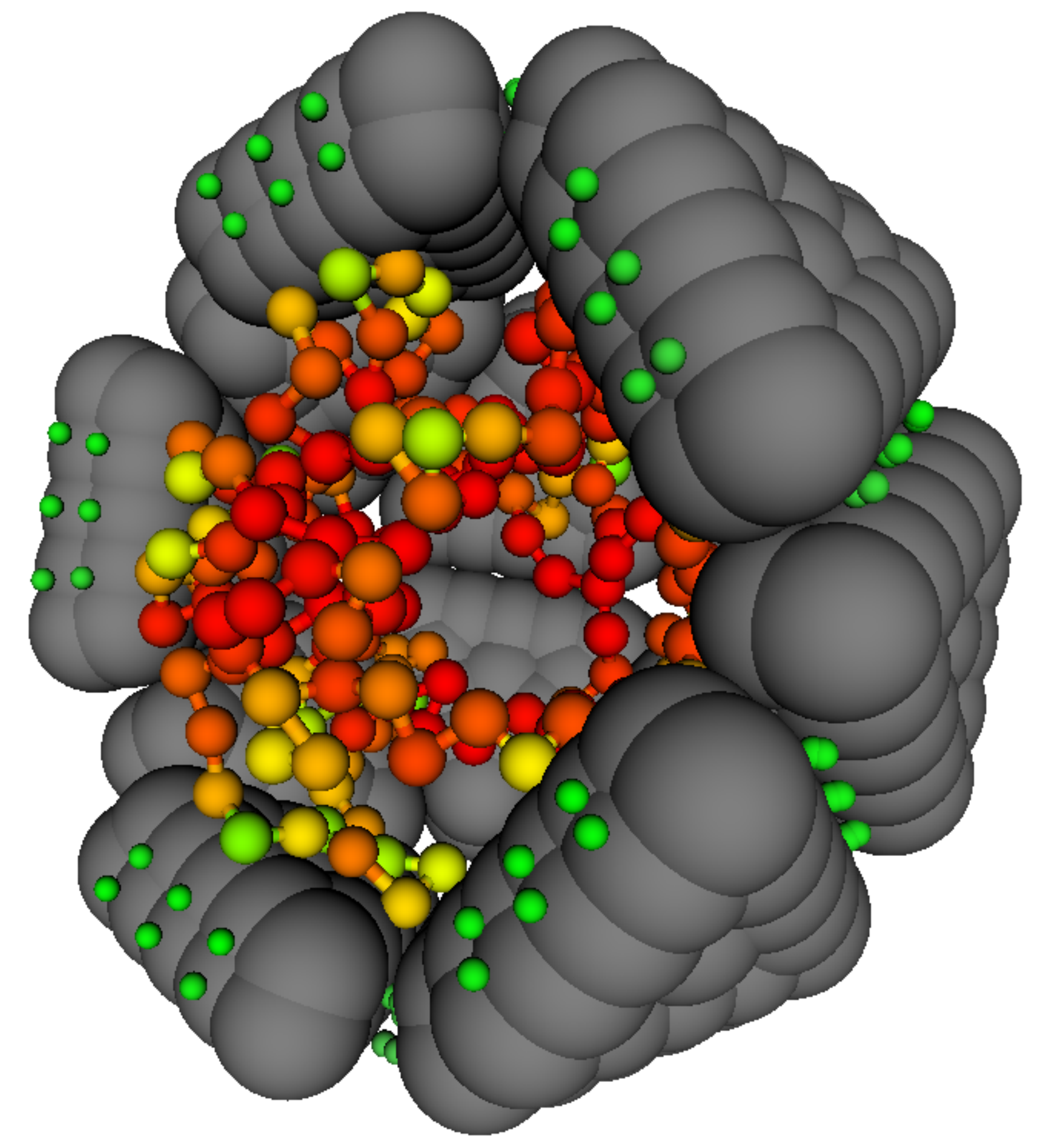}
    \label{fig:zCutPoly}
}
\caption{ {\bf (a) } Image of a trimer of dimers of the MS2 coat protein \cite{Valegard1997}, which was generated from the crystal structure PDBID:1ZDH\cite{Valegard1997} using VMD \cite{Humphrey1996}. The three proteins of the crystal structure asymmetric unit are shown along with the three symmetry-related subunits that complete the dimer subunits. The protein atoms are shown in van der Waals representation, RNA-stem loops are drawn in cartoon format and colored green, and positive charges on the proteins are colored blue. {\bf (b)} The arrangement of polymer attractors on the model capsid subunit, as viewed from inside the capsid. The capsomer-polymer attractors are colored blue and the capsomer-capsomer attractors are colored green. { \bf (c) } A cutaway view of a snapshot of a polymer with $\Lp=200$ segments encapsulated in a well-formed model capsid. \change{Polymer subunits and capsomer-attractors are colored according to their interaction energy: red for non-interacting, green for optimal interaction and a gradient for intermediate states.}
}
\end{center}
\end{figure}

\ssect{Polymer model} We represent the polymer as a freely jointed chain of spherical monomers, with excluded volume that includes effects of screened electrostatic repulsions \cite{Hariharan1998a}. In the absence of any capsomer subunits, the model represents a polymer in good solvent, which behaves as a self-avoiding random walk with radius of gyration $R_g = 0.21 \Lp^{3/5} \sigma\sub{b}{}$, with $\sigma\sub{b}{}$ the monomer diameter. We then add short-ranged attractions to spherical attractors on the interior surface of model capsid subunits that qualitatively represent the effects of screened electrostatic interactions between negative charges on the polyelectrolyte or nucleic acid and positive charges on the interior surface of capsid proteins. While these positive charges are found on flexible N-terminal `ARMs' in many ssRNA viruses, our model was particularly motivated by the small RNA bacteriophages (e.g. MS2), in which the RNA \change{or other polyanions interact} with positive charges on the interior capsid surface. \change{ These interactions} have been characterized over the past two decades through a series of crystal structures of MS2 capsids with different sequences of short RNA hairpins (e.g. \cite{Valegard1994,Worm1998,Grahn2001,Valegard1997,Helgstrand2002}) and more recently, cryo-EM images show the genomic RNA inside the MS2 capsid\cite{Toropova2008}. Fig.~\ref{fig:MS2} shows an image of a trimer of dimers of the MS2 coat protein from the crystal structure highlighting the location of positive charges  and RNA binding sites (a dimer is the fundamental subunit for MS2).  Consistent with the overall simplicity of our model, we crudely capture the geometry of those charges by placing the capsid-polymer attractors as shown in Fig.~\ref{fig:modelpolyatt}. Other arrangements and numbers of attractors sites lead to similar results; however, simulated assembly was less effective when distances between attractors sites were incommensurate with the ground state distance between polymer subunits. The comparison with MS2 is only meant to be suggestive, as for computational simplicity we consider a flexible homopolymer and we model a T=1 capsid with trimers as the basic assembly unit, while MS2 has a T=3 capsid and the dimer is the assembly unit \cite{Stockley2007}.

\ssect{Pair Interaction}

In our model, all potentials can be decomposed into pairwise interactions. Potentials involving capsomer subunits further decompose into pairwise interactions between their constituent building blocks -- the excluders and attractors. The potential of capsomer subunit $i$, $U\sub{cap,}{i}$, with position $\bm R_i$, attractor positions $\{\bm a_i\}$ and excluder positions $\{\bm b_i\}$ is the sum of the a capsomer-capsomer part, $U\sub{cc}{}$, and a capsomer-polymer part $U\sub{cp}{}$:
\begin{eqnarray}
\fl
U\sub{cap,}{i} =
        \sum_{\mathrm{cap\ }{j\neq i}} U\sub{cc}{}(\bm R_i, \{\bm b_i\}, \{\bm a_i\}, \bm R_j, \{\bm a_j\}, \{\bm b_j\})
+       \sum_{\mathrm{poly\ }{k}} U\sub{cp}{}(\bm R_i, \{\bm b_i\}, \{\bm a_i\},\bm R_k)
\label{eq:Ucap},
\end{eqnarray}
where the first sum is over all capsomers other than $i$ and the second sum is over all polymer segments. Similarly the potential of a polymer subunit $i$ is the sum of a capsomer-polymer term, $U\sub{cp}{}$, and a polymer-polymer term, $U\sub{pp}{}$:
\begin{eqnarray}
\fl
U\sub{poly,}{i} =   \sum\sub{poly\ }{j\ne i} U\sub{pp}{}( \bm R_i, \bm R_j ) +
                    \sum\sub{cap\ }{k} U\sub{cp}{}(\bm R_i, \bm R_k, \{\bm b_k\}, \{\bm a_k\})
\end{eqnarray}
where $\bm R$, $\{\bm a\}$ and $\{\bm b\}$ are defined as before. The capsomer-capsomer potential $U\sub{cc}{}$ is the sum of a repulsive potential between every pair of excluders and an attractive interaction between complementary attractors:
\begin{eqnarray}
\label{Ucc}
\fl
U\sub{cc}{}(\bm R_i, \{\bm a_i\}, \{\bm b_i\} ,\bm R_j, \{\bm b_j\}, \{\bm a_j\})  &=&
    \sum_{k,l}^{N\sub{b}{}} \LJ{8} \left(
    \left| \bm{R}_i + \bm{b}_i^k - \bm{R}_j - \bm{b}_j^l \right|,
    \ 2^{1/4} \sigma\sub{b}{},
    \ \sigma\sub{b}{} \right)
    \nonumber \\
    &+&
    \sum_{k,l}^{N\sub{a}{}} \chi_{kl} \ecc \LJ{4} \left(
    \left| \bm{R}_i + \bm{a}_i^k - \bm{R}_j - \bm{a}_j^l \right| - 2^{1/2} \sigma\sub{a}{},
    \ 4\sigma\sub{a}{},
    \ \sigma\sub{a}{} \right)
\end{eqnarray}
where $\ecc$ is an adjustable parameter setting the strength of the capsomer-capsomer attraction \change{ at each attractor site}, $N\sub{b}{}$ and $N\sub{a}{}$ are the number of excluders and attractors respectively, $\sigma\sub{b}{}$ and $\sigma\sub{a}{}$ are the diameters of the excluders and attractors, which are set to 1.0 and 0.20 respectively throughout this work, $\bm{b}_i^k$ ($\bm{a}_i^k)$ is the body-centered location of the $k^\mathrm{th}$ excluder (attractor) on the $i\mathrm{th}$ subunit, $\chi_{kl}$ is 1 if attractors $k$ and $l$ are overlapping in a completed capsid (Figure \ref{fig:modelcapsid}) and 0 otherwise. The function $\LJ{p}$ is defined as a truncated Lennard-Jones-like potential:
\begin{equation}
\label{Leight}
\LJ{p}(x,x_\mathrm{c},\sigma) \equiv
\left\{  \begin{array}{ll}
      \frac{1}{4}\left( \left(\frac{x}{\sigma}\right)^{-p} - \left(\frac{x}{\sigma}\right)^{-p/2} \right) & : x < x_\mathrm{c} \\
      0 & : \mathrm{otherwise}
      \end{array} \right.
      \label{eq:LJ}
\end{equation}
The capsomer-polymer interaction is defined identically to the capsomer attractor potential. For capsomer $i$ with position $\bm R_i$, attractor positions $\{\bm a_i\}$, excluder positions $\{\bm b_i\}$ and polymer subunit $j$ with position $\bm R_j$, the potential is:
\begin{eqnarray}
\fl
U\sub{cp}{}(\bm R_i, \{\bm b_i\}, \{\bm a_i\}, \bm R_k) &=&
    \sum_{k}^{N\sub{b}{}} \LJ{8} \left(
    | \bm R_i + \bm{b}_i^k - \bm R_j |,
    2^{1/4} \sigma\sub{bp}{},
    \sigma\sub{bp}{} \right) \nonumber \\
    &+&
    \sum_{k}^{N\sub{a}{}} \xi_k \ecp \LJ{8} \left(
    | \bm R_i + \bm{a}_i^k - \bm R_j | + 2^{1/4} \sigma\sub{p}{},
    4 \sigma\sub{p}{},
    \sigma\sub{p}{} \right) \\ \nonumber
\sigma\sub{bp}{} &\equiv& \frac{1}{2} \left( \sigma\sub{b}{} + \sigma\sub{p}{} \right)
\end{eqnarray}
where $\ecp$ is an adjustable parameter setting the strength of the capsomer-polymer attraction \change{at each attractor site}, $\sigma\sub{p}{}$ is the diameter of a polymer subunit which is set to $0.4\sigma\rsub{b}$ throughout this work and $\xi_k$ is 1 if attractor $k$ is one of the three central polymer attractors on the subunit (see figure \ref{fig:modelpolyatt}), $1/2$ if $k$ is one of the three outermost polymer attractors and 0 otherwise. The factor of $1/2$ for the outer polymer attractor compensates for the fact that in the ground state of the capsid, each such attractor will overlap with an outer attractor from across the capsomer-capsomer interface. Finally, the polymer-polymer subunit interaction is broken into bonded and non-bonded components, where the bonded interactions are only evaluated for monomers occupying adjacent positions along the polymer chain:
\begin{eqnarray}
\fl
\label{Upp}
U\sub{pp}{}(\bm R_i, \bm R_j) = \left\{  \begin{array}{ll}
                              \LJ{8}(R_{ij}, 2^{1/4}\sigma\sub{p}{}, \sigma\sub{p}{})           & : R_{ij} < 2^{1/4} \sigma\sub{p}{} \\
                              \LJ{8}(2^{5/4}\sigma\sub{p}{} - R_{ij}, 2^{5/4}\sigma\sub{p}{}, \sigma\sub{p}{}) & : R_{ij} > 2^{1/4} \sigma\sub{p}{} \ \& \ \{i,j\}\ \mathrm{bonded} \\
                              0                                                 & : R_{ij} > 2^{1/4} \sigma\sub{p}{} \ \& \ \{i,j\}\ \mathrm{nonbonded}
                              \end{array}
                        \right.
\end{eqnarray}
where $R_{ij} \equiv | \bm R_i - \bm R_j |$ is the center-to-center distance between the polymer subunits.

\ssect{Length Scales}
\label{sec:lengthScales}
Based on the size of a typical T=1 capsid we can assign a value to the simulation unit of length $\sigma\rsub{b}$. Choosing satellite tobacco mosaic virus with outer radius 9.1 nm \cite{Reddy2001}  gives $\sigma\rsub{b} \sim 2.36$ nm and the edge length of our triangular subunits as $\sim 7$ nm \change{and $\sigma\rsub{a}=0.2\sigma\rsub{b}\sim 0.5$nm as the range of the individual capsomer-capsomer attractors. One polymer segment, with diameter $\sigma\rsub{p}=0.4\sigma\rsub{b}$, could represents about 3 base pairs of \change{homopolymeric} ssRNA and our statistical segment length is 1.5 times that of ssRNA. Finally, we will present subunit bath concentrations as $c_0$ with units $\sigma\rsub{b}^{-3}$; the approximate experimental concentration corresponding to our simulations is thus $c_\mathrm{exp} \sim 1.25 \times 10^5 c_0\ \mu\mathrm{M}$, according to which we sample from concentrations of 80 to 500 $\mu\mathrm{M}$. It is important to note, however, that results from this highly simplified model should only be taken to be qualitative and that these length scales, in particular the mapping to concentration, merely serve to identify orders of magnitude.}

\ssect{Dynamics simulations}

We evolve particle positions and orientations from  random non-overlapping initial positions with over-damped Brownian dynamics using a second order predictor-corrector algorithm\cite{Branka1999, Heyes2000}. The capsomer subunits have anisotropic translational and rotational diffusion constants calculated using Hydrosub7.C\cite{Hydrosub}. To represent an experiment with excess capsid protein, the system is coupled to a bulk solution with concentration $c_0$ by performing grand canonical Monte Carlo moves in which subunits more than $10\sigma\rsub{b}$ from the polymer are exchanged with a reservoir at fixed chemical potential with a  frequency consistent with the diffusion limited rate\cite{Hagan2008}. While it is beyond the scope of this manuscript to consider other protein-polymer stoichiometries, the effect of stoichiometry on polymer encapsulation is analyzed with an equilibrium theory in Ref.~\cite{Zandi2009}, and the effects of stoichiometry on the equilibrium and kinetics of the encapsulation of nanoparticles is discussed in Ref.~\cite{Hagan2009}. To mimic a bulk system periodic boundary conditions are employed with the box side length $40\sigma\rsub{b}$.

\ssect{Equilibrium calculation of the driving force for polymer encapsulation.}
To determine the thermodynamic driving force for encapsulation of the polymer in this model, we compute the difference in chemical potential between a free polymer and a polymer encapsulated in a perfect capsid. By computing this chemical potential difference as a function of polymer length, we identify the polymer length that is thermodynamically optimal for packaging. Specifically, we implemented an off-lattice version of the procedure outlined by Kumar et al. \cite{Kumar1991} for calculating the residual chemical potential $\mu\rsub r$ of a polymeric chain:
\begin{eqnarray} \nonumber
-\beta \mu\rsub r(\Lp) &\equiv& -\beta \left\{\mu_\mathrm{chain}(\Lp+1) - \mu_\mathrm{chain}(\Lp)\right\} \\
                     &=& \log \langle \exp(-\beta U_\mathrm{I} (\Lp))\rangle
\label{mur}
\end{eqnarray}
where $\Lp$ is the number of segments in the chain and $U_\mathrm{I}$ is the interaction energy experienced by a test (ghost) segment added to either end of the chain with a random position. The angle brackets in equation~\ref{mur} refer to an equilibrium average over configurations of the chain with $\Lp$ segments and positions of the test segment. Due to the potential between bonded polymer subunits, equation (\ref{Upp}), importance sampling was required for the average to be computationally feasible. The positions of the inserted particles were chosen such that the distance from the test particle to its bonded partner on the chain is drawn from a normal distribution with mean $2^{1/4} \sigma\sub{p}{}$ and standard deviation $0.25 \sigma\sub{p}{}$, truncated at $0.75\sigma\sub{p}{}$. The effect of the biased insertion locations was removed a posteriori according to the standard formula for non-Boltzmann sampling\cite{Frenkel2002,Chandler1987}.

Once the calculation of the average test particle energy was completed for a particular value of $\Lp$, the polymer length was increased by one segment and the calculation was repeated. At each value of $\Lp$, $10^8$ test insertions were performed interleaved with $10^5$ dynamics steps for 50 independent trials. Each calculation began at $\Lp=1$. To calculate the difference in chemical potentials between free and encapsulated polymers, the procedure was performed for an isolated polymer as well as polymers inside capsids. For the latter calculations, the polymer subunit was started inside a well-formed empty capsid. To enhance computational feasibility, the capsid subunit positions were not relaxed during the calculation.

\section{Results}
\label{sec:results}
To understand the influence of polymer properties on capsid assembly, we performed simulations for a range of polymer lengths $\lp$, polymer-subunit interaction energies $\ecp$, capsid subunit-subunit binding energies $\ecc$, and free subunit concentrations $\conc$. The parameters $\ecc$ and $\ecp$ could be experimentally controlled by varying solution $p$H or ionic strength \cite{Ceres2002,Kegel2004}.

\ssect{ Kinetic Phase Diagram }

\label{sec:kpd}

We begin by considering assembly outcomes at the observation time $\tobs=2 \times 10^4 t_0$, which is long enough that assembly outcomes do not vary significantly with time except at short polymer lengths, but is not sufficient to equilibrate kinetic traps if there are large activation barriers. Results are shown for $\logc=-7.38$, which maps to $\sim80$ $\mu$M (see section \ref{sec:lengthScales}) and $\ecc = 4.0 \kt$.  Recalling that $\ecc$ is the energy per attractor this value may seem like a large binding energy, but the short-ranged and stereospecific subunit-subunit interactions involve a large entropy penalty \cite{Hagan2006,Erickson1981,Ben-tal2000}, and dimerization is unfavorable free energetically, with a dissociation constant $\Kd=1$ mM (see \ref{sec:freeenergy}). A rough estimate of the free energy per subunit in a complete capsid for this binding energy is $g_\mathrm{capsid}\approx-9.2\kt$. Spontaneous assembly of empty capsids at this subunit concentration requires $\ecc \gtrsim 5.0 \kt$ or free energy per subunit $g_\mathrm{capsid}\approx-14.5 \kt$, which is consistent with experimental values at which empty capsids assemble ( e.g. \cite{Ceres2002,Parent2006,Zlotnick2003a}).

\begin{figure}
\subfloat[]{
    \includegraphics[width=0.6 \textwidth]{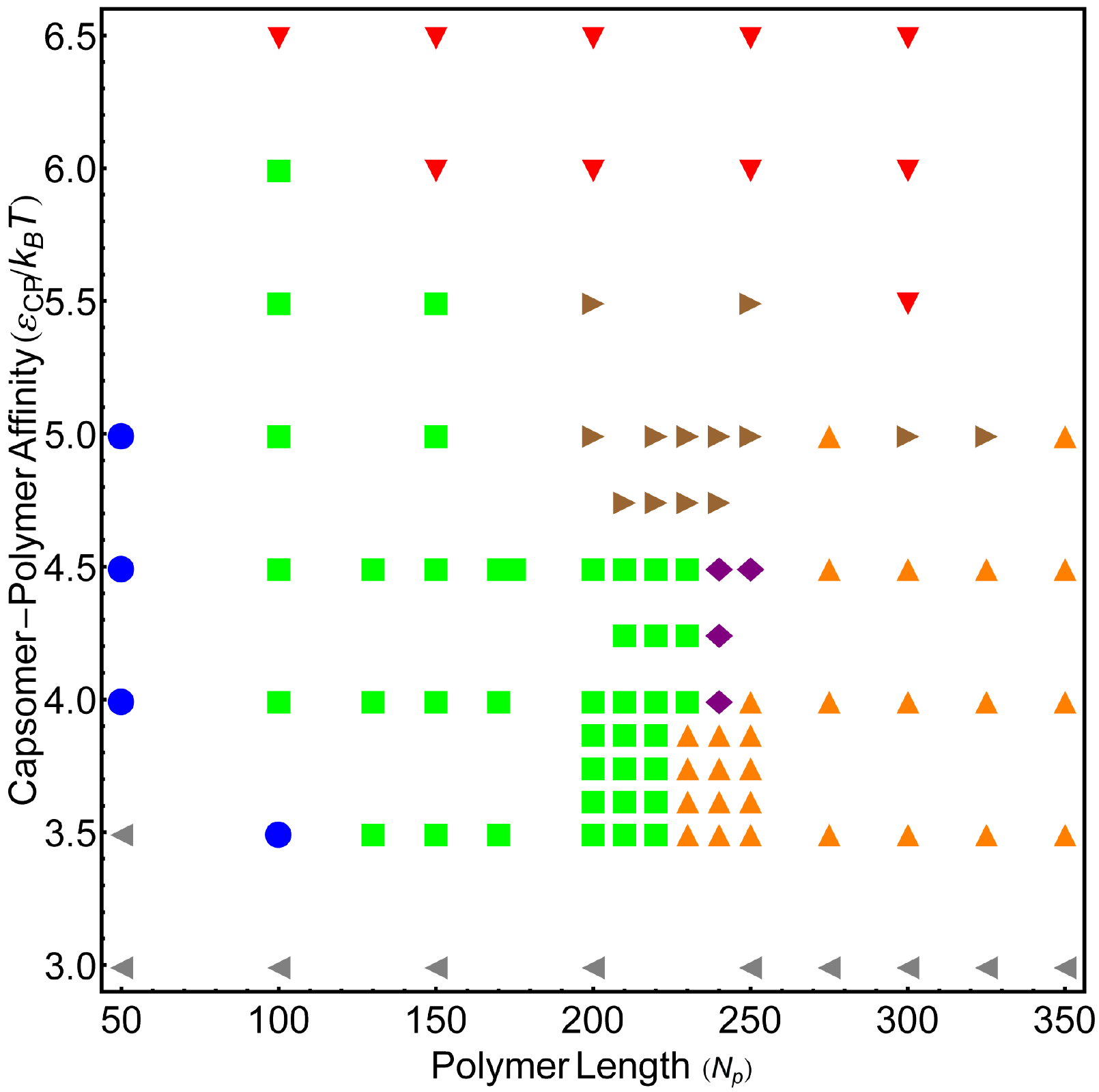}
    \label{fig:kpd}
}
\subfloat[]{
    \includegraphics[width=0.265 \textwidth]{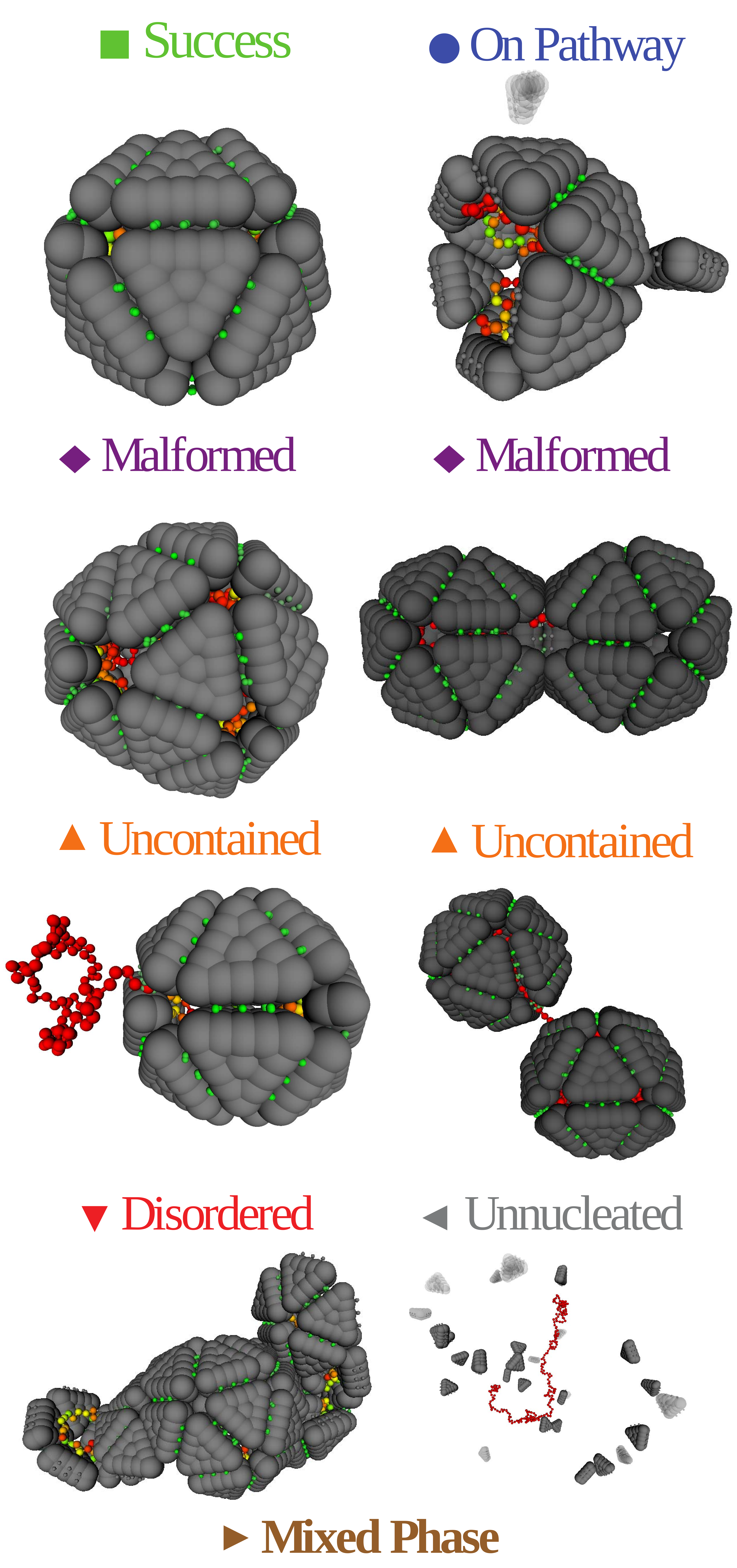}
    \label{fig:kpdleg}
}
\caption{Kinetic phase diagram showing the dominant assembly product as a function of $\lp$ and $\ecp$ for $\ecc$ = 4.0 and $\logc$ = -7.38 at observation time $\tobs=2 \times 10^4 t_0$. The legend on the right shows snapshots from simulations that typify each dominant configuration. Data points indicate the majority outcome, except for the `malformed' and `mixture' points. For malformed points there was a plurality of malformed capsids and a majority of malformed plus well-formed capsids. For points labeled `mixed phase' there was no clear plurality. The exact proportions of the outcomes are available in figure \ref{fig:cumulative}, \ref{sec:further}. Data points correspond to 20 independent assembly trajectories.}
\end{figure}

Fig.~\ref{fig:kpd} is a `kinetic phase diagram', showing the dominant assembly outcome as a function of $\lp$ and $\ecp$ (figure \ref{fig:kpdleg}) at $\tobs$. There is a single region of polymer lengths and interaction strengths in which most polymers are completely encapsulated in well-formed capsids (defined in section~\ref{sec:model} and figure \ref{fig:modelcapsid}). For the remainder of this article, we will refer to complete encapsulation in a well-formed capsid as `successful' assembly. Within this region there are optimal polymer lengths and values of $\epc$ for which the fraction of trajectories ending in success is nearly 100\% (figure \ref{fig:cumulative}, \ref{sec:further}). Notably, polymers that are much larger than the capsid before packaging are successfully encapsulated: the effective capsid inner radius is $2.33 \sigma_\mathrm{b}$, while high success fractions are found for  $\Lp=230$ with unpackaged radius of gyration $\Rg=\change{5.49 \sigma_\mathrm{b}}$ and the longest successfully packaged polymer had $\Lp=300$ and $\Rg=\change{6.43 \sigma_\mathrm{b}}$. This result is  consistent with the experimental observation that polystyrene sulfonate molecules with radii of gyration larger than capsid size were encapsulated in cowpea chlorotic mottle virus capsids \cite{Bancroft1969,Hu2008}.

As the polymer length or $\ecp$ deviate from their optimal values, successful encapsulation yields are reduced by several failure modes. Polymers that are short enough to become completely adsorbed before the capsid finishes assembling tend to result in incomplete, but well-formed `on-pathway' capsids for moderate binding energies $\ecc$. As discussed below, assembly slows dramatically after the polymer is completely encapsulated because the polymer plays both thermodynamic and kinetic roles in enhancing assembly kinetics. \change{We note that if the assumption of infinite dilution of polymers is relaxed, capsids could assemble around multiple short polymers.}

As $\ecp$ or $\lp$ are increased past their optimal values several forms of thermodynamic or kinetic traps hinder encapsulation, hence, weaker subunit-polymer interactions enable packaging of longer polymers. There is a similar nonmonotonic dependence of encapsulation yields with respect to binding energies $\ecc$ or the free subunit concentration (Fig.~\ref{fig:nanl}a below). These observations are consistent with the results of Kivenson et al. \cite{Kivenson2010}, suggesting that the dependence of assembly outcomes on system parameters does not depend strongly on subunit or capsid geometries. However, the present model enables us to examine the morphologies of failure modes as a function of system parameter values and in the presence of correlated polymer-subunit motions. The off-pathway failure modes can be roughly separated into three categories, illustrated by representative snapshots in figure \ref{fig:kpdleg}: (1) Uncontained, in which the capsid closes around an incompletely encapsulated polymer. As discussed in Kivenson et al. \cite{Kivenson2010}, uncontained configurations form when the addition of capsomer subunits and eventual capsid closure is fast compared to polymer incorporation; a large activation barrier hinders complete encapsulation of such configurations. Beyond a certain polymer length, uncontained configurations become thermodynamically favorable (see below). \change{If the polymer is longer still ($\lp \gtrsim 300$), the uncontained segment acts much like a free polymer and nucleates the assembly of a second completed capsid which results in a `doublet', as shown in figure \ref{fig:kpdleg}. For the larger values of $\ecp$ in the uncontained regime, both capsids can nucleate and grow simultaneously. Even longer polymer lengths can lead to multiplets with more than two capsids, similar to structures recently seen in electron microscopy images of cowpea chlorotic mottle virus (CCMV) proteins assembled around RNA molecules with lengths that are multiples of the CCMV genome length\cite{KnoblerPC2010}.} (2) Multiple large partial capsids. When multiple capsids nucleate on the same polymer and grow to significant size ($\sim10$ or more subunits) without associating, they are rarely geometrically compatible for fusion. Even though adsorbed oligomers contact each other frequently due to polymer motions, successful merging from such a configuration is rare because it requires significant subunit dissociation. (3) Defective but closed capsids, which we refer to as `malformed' in this work. For many combinations of large $\Lp$ and $\ecp$ we observe closed shells with \change{hexameric dislocations (figure \ref{fig:hexamers}) that resemble the closed structures found by Nguyen et al. \cite{Nguyen2009} for T=1 capsids, noting that we only consider trimeric subunits here. We also find structures in which two well-formed capsids share a single triangular face (see figure \ref{fig:kpdleg}), reminiscent of the structure of many geminiviruses\cite{Zhang2001}.}

\ssect{Comparison to equilibrium results.}  Since the assembly outcomes in figure \ref{fig:kpd} are measured at finite observation times, they identify configurations that are metastable on assembly time scales, and therefore relevant to in vitro experiments and viral replication in vivo. To fully understand the relationship between driving forces and assembly yields, it is interesting to compare these results to equilibrium thermodynamics. We therefore measured the chemical potential for a polymer encapsulated in a well-formed capsid $\mu_\mathrm{chain}^\mathrm{cap}$ and that for a free polymer $\mu_\mathrm{chain}$(see section~\ref{sec:model}). The difference $\mu_\mathrm{chain}^\mathrm{cap} - \mu_\mathrm{chain}$ measures the equilibrium driving force to completely enclose the polymer in a well-formed capsid, and is a typical result of an equilibrium calculation (e.g. \cite{Siber2008,Harvey2009,Belyi2006,Schoot2005,Jiang2009}).

\begin{figure}[!h]
\begin{center}
\subfloat[]{
\includegraphics[width=0.5 \textwidth]{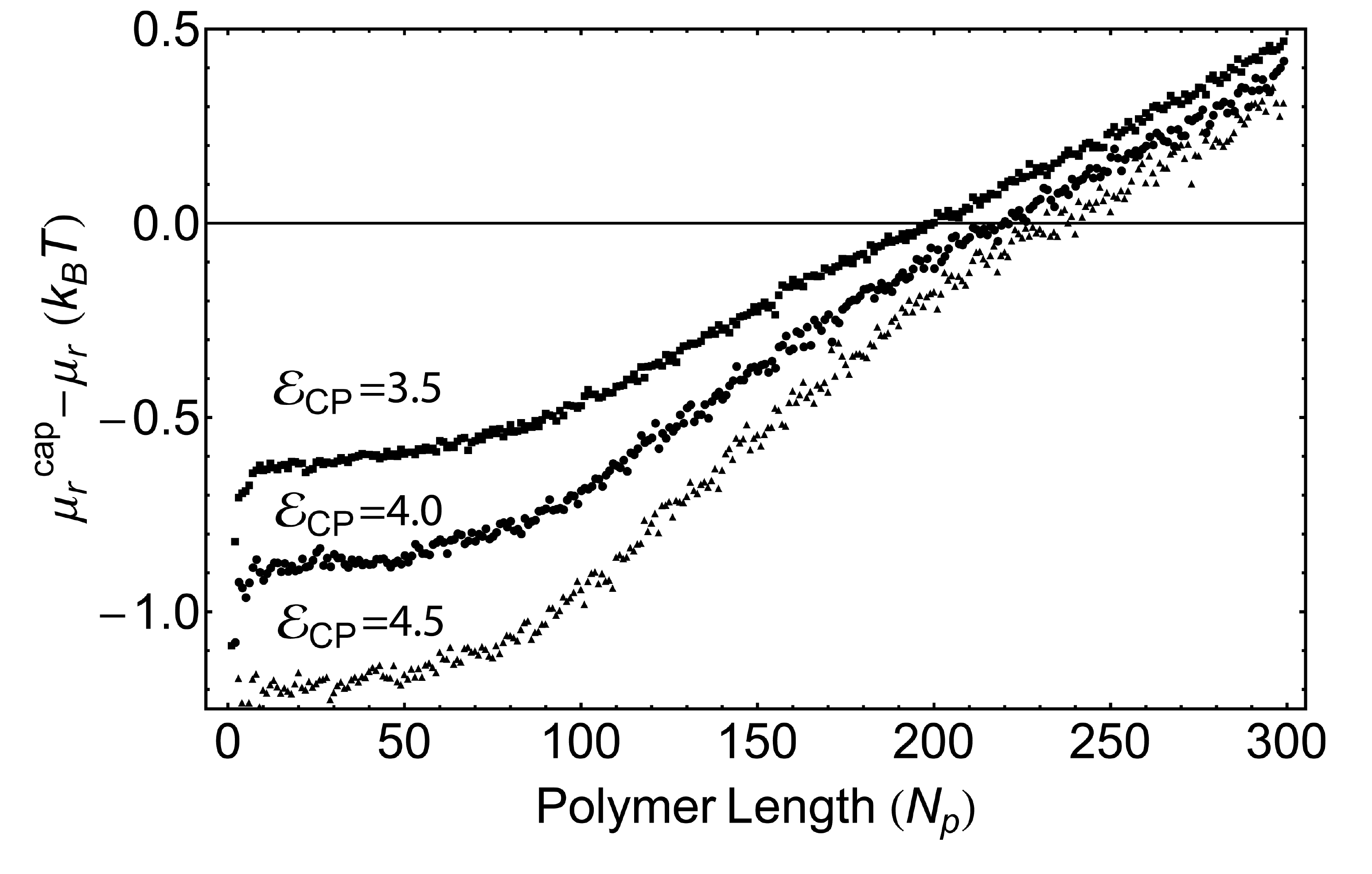}
\label{fig:rcpplot}
}
\subfloat[]{
\includegraphics[width=0.5 \textwidth]{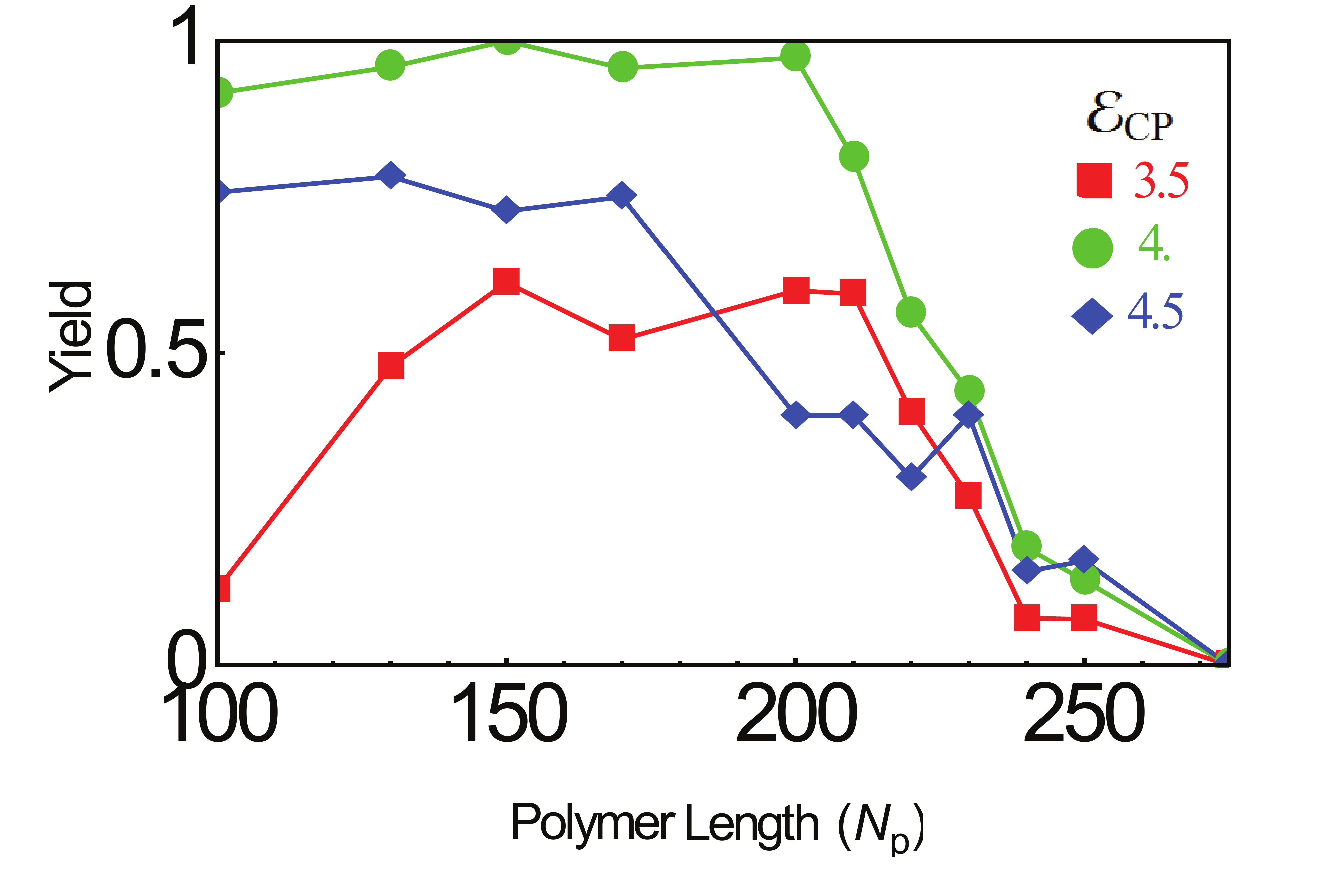}
\label{fig:rcpsuc}
}
  \caption{ {\bf a)} Residual chemical potential difference between a polymer grown inside a well-formed capsid and a free chain, $\mu_\mathrm{chain}^\mathrm{cap} - \mu_\mathrm{chain}$, at indicated capsomer-polymer affinities $\ecp$. {\bf b)} The fraction of Brownian dynamics trajectories that end with a polymer completely encapsulated in a well-formed capsid is shown for the same capsomer-polymer affinities.  }
  \label{fig:rcp}
\end{center}
\end{figure}

The residual chemical potential difference $\mu\rsub r^\mathrm{cap}-\mu\rsub r$, which gives the change in driving force upon increasing the polymer by a single segment, is shown for several values of $\ecp$ in figure \ref{fig:rcpplot}. The thermodynamically optimal polymer length for packaging in a well-formed capsid, $\Lpeq$, corresponds to the length at which $\mu\rsub r^\mathrm{cap}-\mu\rsub r = 0$. In contrast to the kinetic results described above, we see that $\Lpeq$ monotonically increases with $\ecp$:  $\Lpeq \approx$ 195, 220, 230 for $\ecp=3.5$, 4.0, 4.5 respectively. For comparison, the fraction of successful dynamical assembly trajectories is shown as a function of $\Lp$ in  figure \ref{fig:rcpsuc}, where we see that the highest yields are obtained for the intermediate $\ecp=4.0$. All values of $\ecp$ show a sharp decrease in yields of well-formed capsids as the polymer length approaches $225 \lesssim \Lp \lesssim 250$; the drop-off point is nearly insensitive to $\ecp$ (although still nonmonotonic). Interestingly, while this polymer length is close to the thermodynamically optimal polymer lengths it does not reproduce their dependence on $\ecp$.

As shown in Figs. \ref{fig:kpd} and \ref{fig:cumulative}, the uncontained failure mode is largely responsible for the sharp drop-off in well-formed capsid yields at large polymer lengths. While uncontainment can occur out-of-equilibrium if the capsid closes faster than the polymer is incorporated, it becomes thermodynamically favored over a well formed capsid above a particular polymer length. The `uncontainable length' $\Lpmax$ can be estimated as follows. The residual chemical potential difference $\mu\rsub r^\mathrm{cap}-\mu\rsub r$ is roughly 0 for the uncontained portion of the polymer, so the lowest free energy figure configuration of an uncontained polymer would have the thermodynamically optimal length contained and the remainder uncontained. The uncontained configuration becomes thermodynamically favored over a well-formed capsid when the integrated residual chemical potential difference becomes larger than the capsomer-capsomer strain free energy in an uncontained configuration. The strain energy  was measured in the simulations to be $\sim 10-20\kt$.  Neglecting capsomer entropy differences between well formed and uncontained configurations, comparison of this value with  figure~\ref{fig:rcp} estimates that uncontained polymers become thermodynamically favored at $\Lpmax\approx 250$ for $\ecp=3.5$, which is close to the drop-off length.  Above this length simulation results show predominantly  uncontained polymers (figure \ref{fig:kpd}).

From figure \ref{fig:stack4.5} we can also see that with strong interactions ($\ecp >= 4.5$), there is a rise in the production of malformed capsids -- larger closed structures containing hexameric dislocations (as in figure \ref{fig:hexamers}). For longer polymers, these defective structures compete thermodynamically with well-formed and/or uncontained configurations since they permit more capsomer-polymer contacts while incurring about  $12-17 \kt$ of strain energy. Their prevalence even at moderate polymer lengths, by contrast, is a kinetic effect that results from the strong capsomer-polymer interactions preventing the defects from annealing.
\change{As discussed for empty capsid assembly in Refs. \cite{Hagan2006,Jack2007,Rapaport2008,Hagan2010}, kinetic traps dominate in an assembly reaction when the time to add new subunits is short compared to the time required for partial capsids to anneal defects or `locally equilibrate'. Annealing requires the disruption of favorable but imperfect interactions, and frequently occurs through the dissociation of improperly bound subunits (as discussed further in section \ref{sec:mechanisms}). The annealing time therefore increases exponentially with $\eb$ and $\epc$, while the subunit association time decreases with $\conc$ or $\epc$ (section  \ref{sec:mechanisms})).} Thus results at a finite observation time deviate more strongly from equilibrium as any of these parameters is increased. A comparison of the kinetic results to an equilibrium calculation that considers all possible assembly products is desirable but beyond the scope of this work.

{\bf Comparison to experimental lengths.} Based on the length scales assigned in section \ref{sec:lengthScales}, the optimal and maximal polymer lengths correspond to approximately 500-750 nucleotides, which is shorter than the 1000 nucleotide genome length of STMV. \mfhch{The optimal length could have been adjusted by adding additional attractor sites--simulation results suggest that the optimal polymer length is roughly linear in the number of attractors in the regime that we have considered, although it depends on attractor spacing and eventually saturates. At this level of simplification there is not an exact mapping between number of charges on capsid proteins and the number of attractor sites, especially considering the complexities associated with changes in the amount of counterion condensation that occur when charged polymers adsorb onto charged capsid proteins.  However, we did not adjust the number of attractors because the results do not change qualitatively, and we did not aim for quantitative accuracy from such a simplified model that does not explicitly calculate electrostatics. Finally, the optimal length might also change if flexible ARMs \cite{Siber2008} and/or representations of base-pairing that lead to compact structures\cite{Zandi2009,Kivenson2010} are considered.}

\subsection{Assembly Mechanisms}

\label{sec:mechanisms}

\begin{figure}[!h]
\begin{center}
\subfloat[]{
\includegraphics[width=0.45 \textwidth]{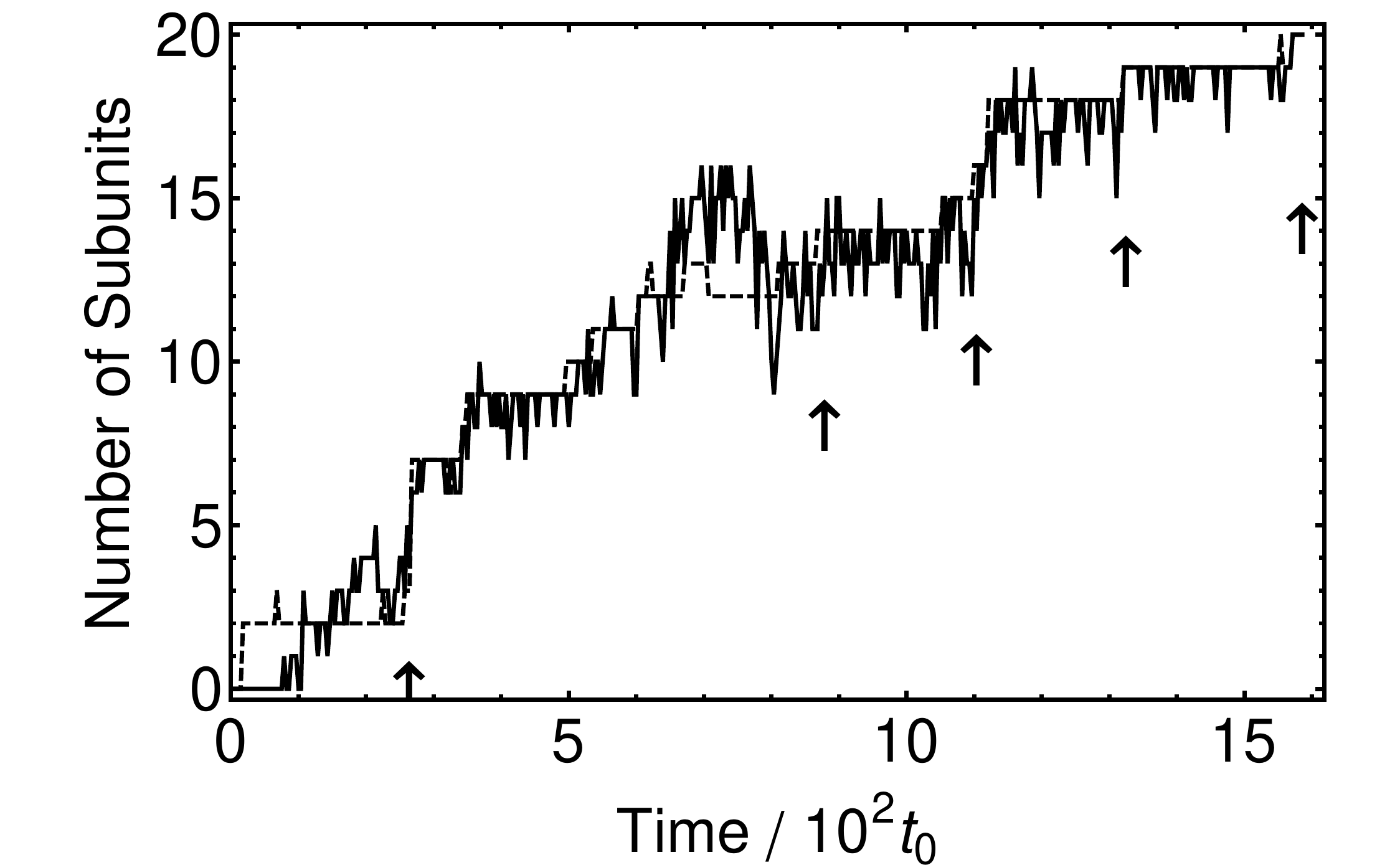}
\label{fig:nanltracelow}
}
\subfloat[]{
\includegraphics[width=0.45 \textwidth]{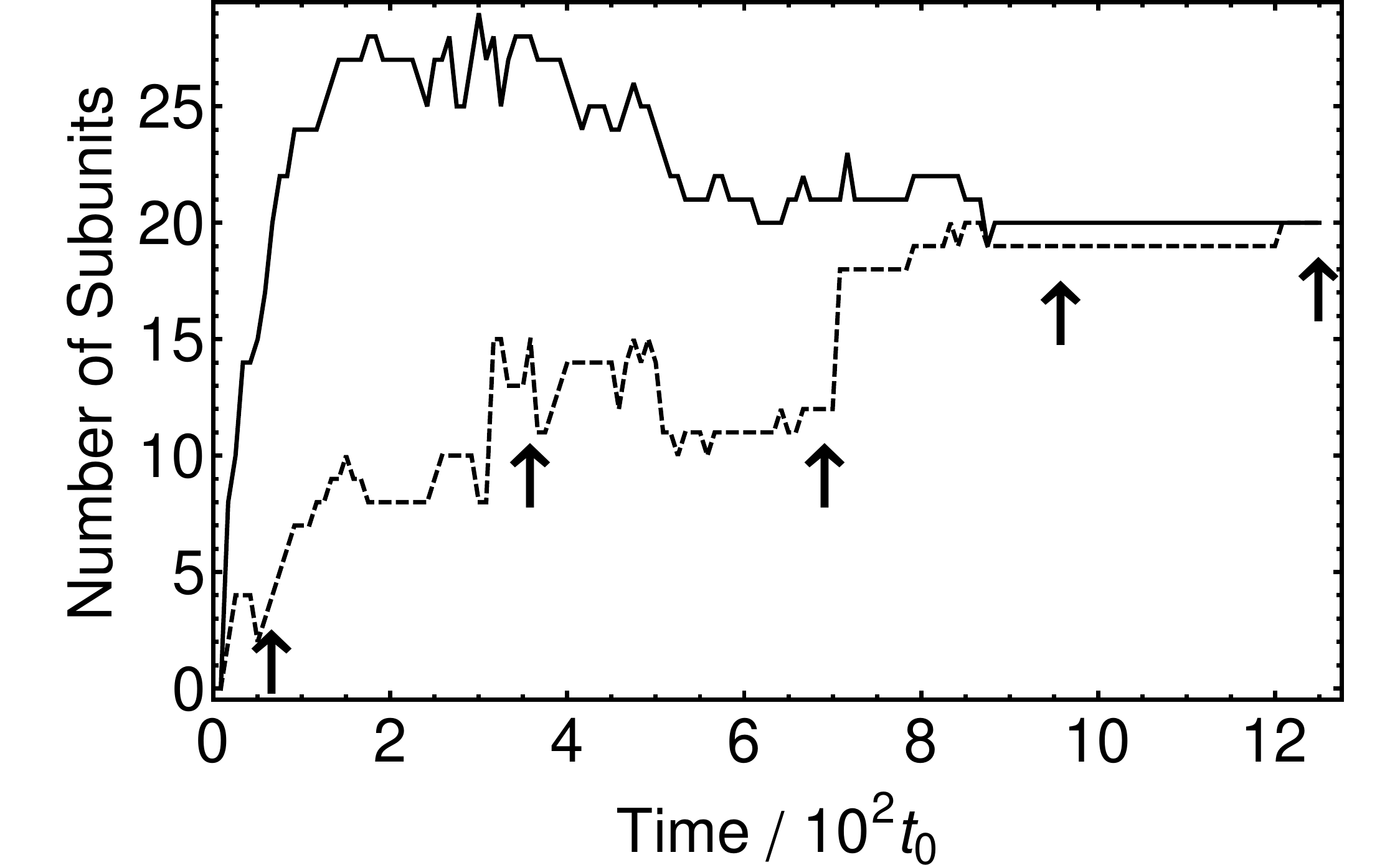}
\label{fig:nanltracehigh}
}
\\
\subfloat[]{
\includegraphics[width=0.95 \textwidth]{Figures/Fig5C}
\label{fig:nanlsnaplow}
}
\\
\subfloat[]{
\includegraphics[width=0.95 \textwidth]{Figures/Fig5D}
\label{fig:nanlsnaphigh}
}
\caption{ Two mechanisms for assembly around the polymer. {\bf (a,b)} The number of capsomer subunits adsorbed onto the polymer (solid) and the size of the largest partial capsid (dashed) are shown as a function of time for {\bf (a)} a trajectory with low $\nanlop$ (the sequential assembly mechanism) and {\bf (b)} a trajectory exhibiting high $\nanlop$ (the en masse mechanism). Parameters are {\bf (a)} $\lp=200$, $\ecp=3.0$, $\logc=-6.5$, $\ecc=4.5$ and {\bf (b)} $\lp=150$, $\ecp=4.5$, $\logc=-5$, $\ecc=3.25$. {\bf (c)} Snapshots from the simulation trajectory shown in {\bf (a)} (points marked with arrows). {\bf (d)} Snapshots corresponding to points marked with arrows in {\bf (b)}  showing the the mass adsorption of subunits onto the polymer followed by annealing of multiple intermediates and finally completion. Once the polymer is completely contained within the partial capsid (second to last frame), addition of the last subunit is relatively slow as discussed in the text.  }
\label{fig:snapshots}
\end{center}
\end{figure}

\begin{figure}[!h]
\begin{center}
\includegraphics[width=0.95 \textwidth]{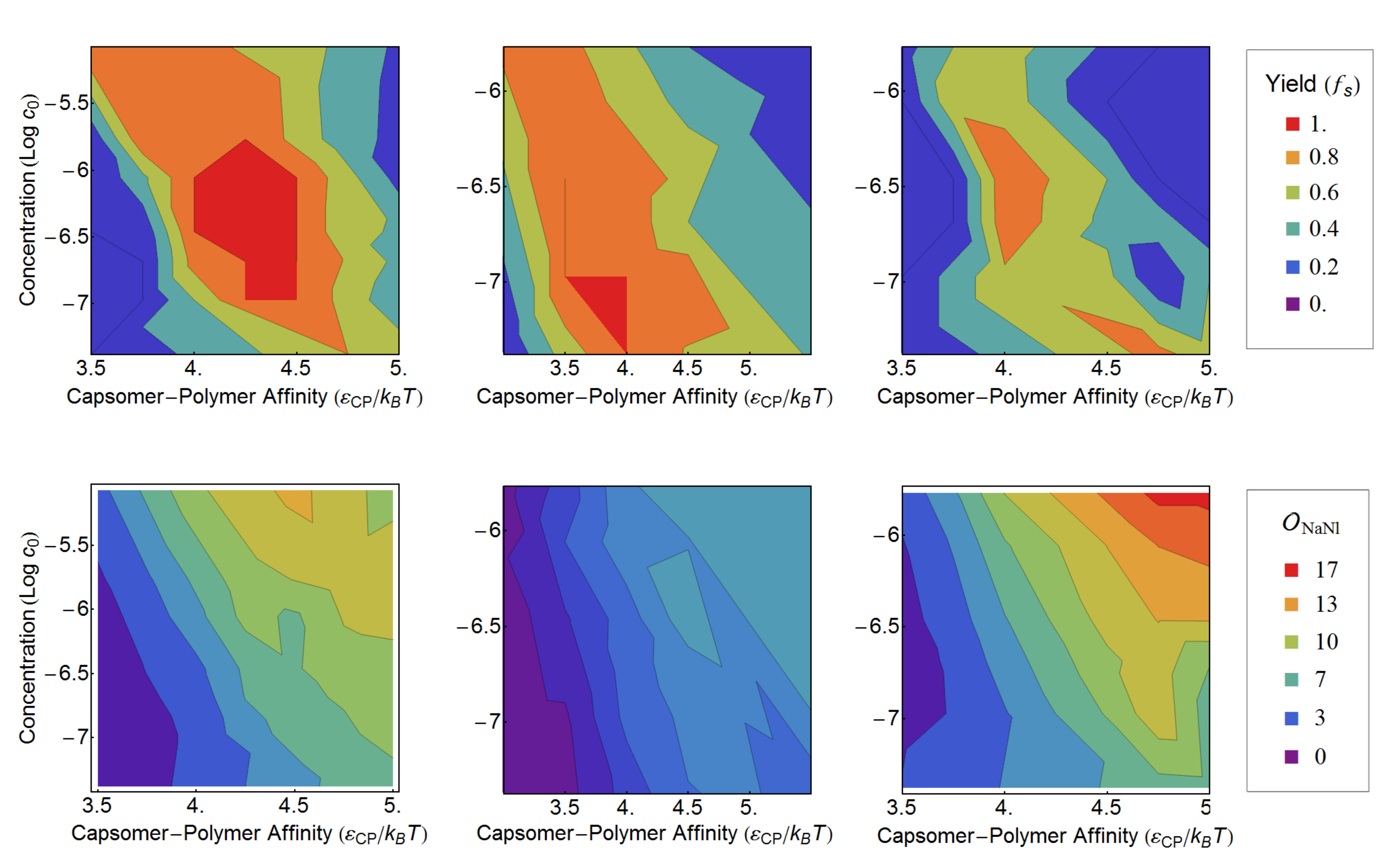}
\caption{ Contour plots of (top panels) the yield, or fraction of trajectories that end with well formed capsids and (bottom panels) the assembly mechanism order parameter $\nanlop$ defined in the text. Plots are shown as functions of $\ecp$ and $\logc$ for parameter values \{$\ecc=3.25, \Lp=150$\} (left), \{$\ecc=4.0, \Lp=150$\} (center), \{$\ecc=3.25, \Lp=200$\} (right). }
\label{fig:nanl}
\end{center}
\end{figure}

In this section we discuss the mechanisms of polymer encapsulation and how these mechanisms depend on the system control parameters. Assembly trajectories can be described by two modes, depending on the rate and free energy for subunits to adsorb to the polymer. Typical trajectories that illustrate each of these modes are shown in  figure \ref{fig:snapshots}. When subunit-polymer association is slow or relatively unfavorable (figure \ref{fig:snapshots}a,c), assembly first requires nucleation of a small partial capsid on the polymer, followed by a growth phase in which one or a few subunits sequentially and reversibly bind to the partial capsid.  Polymer encapsulation proceeds in concert with capsid assembly in this mode. In the alternative mode subunits adsorb on to the polymer en masse in a disordered fashion and then must cooperatively rearrange to form an ordered capsid (figure \ref{fig:nanlsnaphigh}). Assembly occurs rapidly as multiple oligomers appear and coagulate to form an ordered capsid. In the particular trajectory shown, the reordering of subunits results in the polymer contained within a capsid missing one subunit; the final subunit binds after a delay (see discussion of assembly rates below for further discussion). 


To classify trajectories according to these modes, we define an order parameter $\nanlop$, which measures the number of subunits adsorbed onto the polymer that are not in the largest partial capsid, averaged over all recorded snapshots in which the largest assembled partial capsid has a size in the range $3 \leq N_\mathrm{largest} \leq 8$. Large values of the order parameter $\Nanl\gtrsim 8$ indicate that nearly enough subunits to form a capsid have adsorbed before significant assembly occurs (corresponding to the en masse mechanism), while small values $\Nanl \sim 2$ correspond to the sequential assembly mechanism. Values of $\Nanl$ are presented as functions of the system control parameters in figure \ref{fig:nanl} (bottom panels), where we see that the en masse mechanism dominates when subunit adsorption onto the polymer is free energetically favorable and is fast compared to capsid assembly. Specifically, the number of adsorbed subunits approaches or exceeds the number of subunits in a capsid, $c_1 \Lp \gtrsim \Nc$, with $c_1$ with a one-dimensional concentration of adsorbed but unassembled subunits and $\Nc$ the capsid size. In order to reach this limit, the polymer-capsid affinity and free subunit concentration must be large enough that the equilibrium number of adsorbed subunits reaches $\Nc$ even at $\ecc=0$, or $\csurf \Lp \gtrsim \Nc$.  Furthermore, subunit adsorption must approach this equilibrium value faster than the capsid nucleation time $\tnuc$, so that assembly does not deplete $c_1$. Since nucleation times decrease with increasing concentration and binding energy as $\tnuc \sim c_1^{-\nnuc} \exp^{-\eb}$ \cite{Kivenson2010} (see \ref{sec:polynuc}), these conditions are only met for relatively low binding energies $\eb$, as can be seen by comparing figure \ref{fig:nanl} with the values of $\csurf$ shown in figure \ref{fig:csurf}. \change{Furthermore, low binding energies facilitate annealing of imperfect geometries and the desorption of subunits from partial capsids and/or the polymer, which are essential elements of the en masse mechanism. As evident in  Fig.~\ref{fig:nanltracehigh}, it is common for the number of adsorbed subunits to exceed the number in a complete capsid; the excess subunits must unbind before the polymer can be completely encapsulated. Similarly, the en masse mechanism frequently involves the association of large oligomers, which often result in imperfect binding geometries. Annealing of imperfect geometries can occur via rearrangement, but typically involves the dissociation of some subunits.}

To learn how assembly mechanisms correlate to polymer encapsulation efficiency, we also present the fraction of successful assembly trajectories in figure \ref{fig:nanl} (top panels). We first consider the relatively short polymer length $\Lp=150$, for which there are more interaction sites than polymer segments, and the extremely low binding energy $\eb=3.25$ (we did not observe significant yields of assembled capsids with $\eb\leq 3$ for any parameter sets). For these parameters, assembly yields increase with $\csurf$ until high values of $\conc$ and $\ecp$, and significant yields occur only for parameters in which the en masse mechanism dominates. The latter result can be understood by noting that the $\eb=3.25$ corresponds to a large critical nucleus and a large critical subunit concentration and thus no assembly occurs without a high value of $c_1$. In contrast, for $\eb=4$ significant packaging efficiencies are found only when the sequential mechanism dominates. As noted in the previous paragraph, extremely high $\conc$ is required to achieve subunit adsorption rates that are fast compared to assembly time scales at this binding energy. Assembly is not efficient at those concentrations because of kinetic traps.

A similar dependence of packaging efficiencies on $\ecp$ and $\conc$ is found for longer polymer lengths (e.g. $\Lp=200$ in the right panel of figure \ref{fig:nanl}), except that packaging becomes less successful with increasing $\csurf$ in the en masse region even at low $\eb$.  This trend occurs because mass adsorption onto the longer polymer frequently results in multiple nuclei that are unable to simultaneously anneal and encapsulate the polymer and instead yield disordered aggregates, as shown in figure \ref{fig:disordered}.

\ssect{The polymer enhances assembly rates.}

\begin{figure}
\subfloat[] {
  \includegraphics[width=0.5 \textwidth]{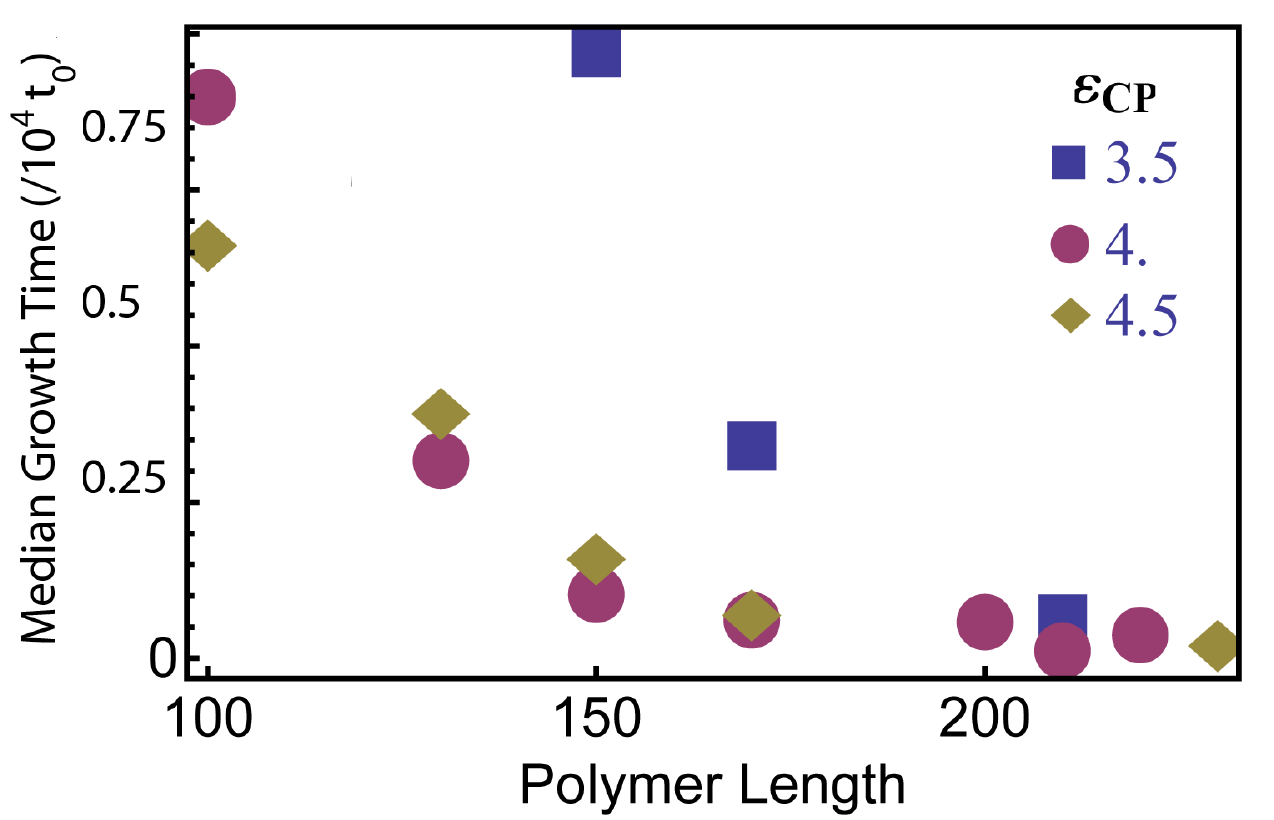}
  \label{fig:growthTimes}
}
\\
\subfloat[] {
    \includegraphics[width=.625\textwidth]{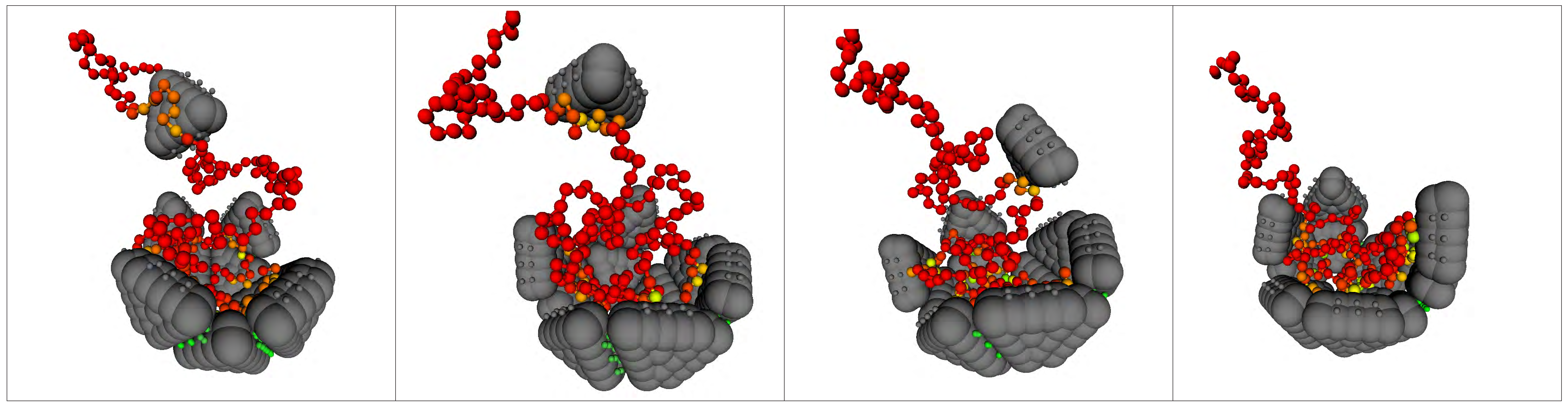}
    \label{fig:slidecookie}
}
\caption{ {\bf a)} The median growth times (time between nucleation and completion) as a function of $\Lp$ for indicated values of $\ecp$, with $\ecc=4.0$ and $\logc=-7.38$. {\bf b)} Snapshots from an assembly trajectory demonstrating both sliding, or one-dimensional diffusion of subunits along the polymer, and the `fly-casting' mechanism described in the text. A free subunit binds the polymer (first frame) and slides towards the growing edge (second and third frame). It then binds to the growing edge of the capsid (fourth frame) while still attached to the polymer, forming a small loop. Note that fly-casting is not limited to such short loops.}
\end{figure}

In addition to affecting assembly outcomes, properties of the polymer have a dramatic effect on assembly timescales. The polymer significantly lowers the free energy barrier for nucleation by stabilizing pre-nucleated partial capsid intermediates, and as discussed next can increase subunit association rates before and after nucleation. The effect of the polymer on nucleation rates is described in \ref{sec:polynuc} and in Ref. \cite{Kivenson2010}.

To quantify the effect of the polymer on rates of growth after nucleation, we measured growth times, or the times between nucleation and completion, for individual capsids. As shown in Fig.~\ref{fig:growthTimes}, the median growth time decreases with polymer length for all interaction parameters until reaching a parameter-independent limiting value at approximately $\Lp=200$.
This trend reflects several mechanisms by which the polymer can influence capsid growth. First, as noted in  \cite{Kivenson2010} binding to the polymer stabilizes partial-capsid intermediates; this is a thermodynamic effect that increases the net rate of assembly by decreasing the rate of subunit desorption from adsorbed intermediates. This effect is particularly important for the conditions we study, where empty capsids do not form spontaneously in the absence of a polymer. Under these conditions assembly slows significantly once the polymer is completely adsorbed in a partial capsid, resulting in the on-pathway incomplete capsids discussed in section \ref{sec:kpd} for short polymers. The effect of increasing  $\epc$ on growth rates saturates when the unbinding rates of polymer-stabilized subunits become small compared to association rates.

The polymer also enhances growth rates by increasing the flux of subunits to and from the assembling partial capsid. Subunit flux is enhanced by (at least) two mechanisms: (1) correlated polymer-subunit motions drag adsorbed subunits to/from binding sites (i.e. the polymer acts like a fly-caster or the Cookie Monster), and (2) adsorbed subunits undergo effectively one-dimensional diffusion (sliding) along the polymer \cite{Hu2007b,Kivenson2010}. While sliding was examined in \cite{Kivenson2010}, correlated polymer-subunit motions were not well represented by the single particle Monte Carlo moves used in that work.  We find that both mechanisms occur in the simulations discussed here; examples can be seen in figure \ref{fig:slidecookie}. For the parameters and model geometries that we use here, correlated polymer-subunit motions are more productive than sliding, and become more important as $c_1$ increases; the en masse assembly mechanism described above is essentially  the extreme limit of correlated polymer-subunit motions at high $c_1$. The flux-enhancement increases with polymer length and $\ecp$ until the rate of transfer of subunits from the polymer to capsid binding sites becomes rate-limiting.

{\bf Completion phase.} The effect of the polymer on subunit association rates leads to a complicated dependence of growth rates on the partial capsid size and system parameters, as illustrated by the two trajectories shown in  figure \ref{fig:snapshots}. In general, net growth rates slow as the partial capsid nears completion because fewer potential binding sites remain available and because the rate at which the polymer captures free subunits diminishes as it is progressively contained.  This trend can be seen in the sequential assembly trajectory shown in figures \ref{fig:nanltracelow} and \ref{fig:nanlsnaplow}. However, because the polymer is relatively long $\Lp=200$ and the capsomer-polymer affinity is relatively weak $\ecp \leq 3.5 $, the polymer makes frequent excursions outside of the partial capsid at all sizes and continues to enhance the subunit flux until the final subunit is in place. In contrast, the trajectory with high $\nanlop$ (figures \ref{fig:nanltracehigh} \& \ref{fig:nanlsnaphigh}) exhibits rapid growth during the rearrangement of adsorbed subunits, but stalls when the polymer becomes completely encapsulated within the capsid missing a single subunit (fourth frame). In this case with a shorter polymer $\Lp=150$ and stronger capsomer-polymer affinity $\ecp=4.5$ the polymer remains completely incorporated and plays no role in attracting the final subunit. As a result, insertion of the final subunit is slow compared to the rest of the assembly process.

We note that the effect of polymer incorporation on the rate of insertion of the last subunit can be significant, since for empty capsid assembly the subunit addition rate decreases somewhat as the capsid nears completion. In our model the last subunit associates on average $\sim$4 times more slowly than those added when the partial capsid is half complete. Unlike the model studied in Nguyen et al.\cite{Nguyen2007}, however, insertion of the final subunit is free energetically favorable, and is not rate limiting under reasonable conditions.

\subsection{Polymer order}

Consistent with experiments (e.g. \cite{Tihova2004,Toropova2008,Schneemann2006}) and the equilibrium calculation of Forrey et al. \cite{Forrey2009} the polymer adopts the symmetry of its capsid, as shown in  figure~\ref{fig:icospoly}. The polymer order arises as a simple consequence of the symmetric arrangement of low free energy sites on the interior capsid surface. To obtain the images in figure ~\ref{fig:icospoly}, we discretized space, and colored each bin with an intensity proportional to the log of the local polymer density $\rho$. In order for the high-density regions to be visible, bins with $\log \rho / \log \rho_\mathrm{max} < 0.25$, with $\rho_\mathrm{max}$ the maximum density, were rendered invisible.

\section{Conclusions}
\label{discussion}

In summary, the calculations in this work show that subunits equipped with interactions driving the formation of an icosahedral shell can assemble into a rich array of structures around a polymer. The nature of the assembly products can be tuned by changing experimentally controllable parameters, such as polymer length, solution conditions, and protein concentrations. Furthermore, the mechanism by which assembly takes place can be systematically varied from a sequential process resembling empty capsid assembly to an en masse process in which subunits rapidly adsorb and then collectively rearrange into an ordered capsid.

The simulations indicate that the en masse mechanism occurs only when the subunit-subunit binding energy is much weaker than that required for empty capsid assembly and there is a strong driving force for subunit absorption onto the polymer. These criteria are met by many single-stranded RNA viruses at physiological conditions, for which protein-protein interactions are too weak to drive empty capsid assembly \cite{Ceres2002} and there are strong electrostatic interactions between the nucleic acid and capsid subunits. In particular, Brome mosaic virions have been described as `loose assemblies' which cannot maintain structural integrity without protein-nucleic acid and protein-divalent cation interactions \cite{Larson2005, Lucas2002, Kaper1975,Johnson1985}. 

Given these observations, it might be surprising that the simulations predict that assembly via the en masse mechanism is less robust than the sequential assembly mechanism, in the sense that high yields of polymers completely encapsulated in well-formed capsids are found over smaller ranges of parameter values (e.g. compare Figs.~\ref{fig:nanl}b and \ref{fig:nanl}d). However, the simulations model assembly around a linear polymer, while secondary and tertiary interactions in RNA molecules lead to compact branched structures \cite{Yoffe2008}. We speculate that polymer compactification due to base pairing could increase the robustness of the en masse mechanism, since it brings the problem closer to the limit of assembly around a rigid core \cite{Hagan2008}.

\ack
The analogy between correlated polymer-capsomer motions and fly-casting was suggested by Charles L. Brooks III, while the Cookie Monster was suggested by Dan Reeves.  MFH and OME were supported by Award Number R01AI080791 from the National Institute Of Allergy And Infectious Diseases.  MFH also acknowledges support by National Science Foundation through the Brandeis Materials Research Science and Engineering Center (MRSEC). Simulations were performed on the Brandeis HPCC.
\vspace{3in}

\bibliographystyle{unsrt}
\bibliography{triangles}

\newpage
\appendix
\section{The effect of the polymer on nucleation times}
\label{sec:polynuc}
To understand the effect of the polymer on nucleation times, we build upon what is known about nucleation times for empty capsid assembly. Several references have analyzed capsid nucleation through simplified rate equations and/or classical nucleation theory \cite{Endres2002,Zandi2006,Hagan2010}, and find that nucleation times can be expressed as $(\tnucempty)^{
-1} \propto f c_0^{n} \exp(G_{n-1}/\kt)$ with $n$ the critical nucleus size, $G_{n-1}$ the interaction free energy of the largest unstable partial capsid (the amount by which subunit-subunit interactions decrease the nucleation barrier), and $f$ a rate constant. Roughly speaking, the concentration of intermediates just below the nucleus size is $c_0^{n-1}\exp(-G_{n-1}/\kt)$ and the rate at which a subunit associates to a pre-nucleus is $f c_0$ (a different attempt rate is derived under the continuum approximation of Ref.~\cite{Zandi2006}). It is important to note that an important simplifying assumption is made in these theories, namely that the identity of a critical nucleus can be defined by the number of subunits alone; i.e., the intermediate size is a sufficient reaction coordinate. This assumption was mildly violated in the simulations of Ref. \cite{Kivenson2010}. Furthermore, for icosahedral capsids it is likely that critical nuclei correspond to particular small polygons (e.g. \cite{Zlotnick2000,Sorger1986,Basnak2010}), and different assembly pathways for a given virus could proceed through critical nuclei with different numbers of subunits\cite{Basnak2010}.

The empty capsid nucleation picture can be extended to include a polymer by noting that adsorption of subunits onto the polymer affects both the free energy barrier and the attempt rate. The concentration of pre-nuclei on the polymer can be expressed as $c_{n-1} \simeq \Lp c_0^{n-1} \exp[-(G_{n-1} + \alpha(\nnuc-1) \gcp)/\kt]$ with $\gcp$ the polymer-subunit interaction free energy (see \ref{sec:freeenergy}), $\alpha$ the fraction of potential polymer-subunit contacts in a typical nucleus, and $G_{n-1}$ the total partial capsid subunit-subunit interaction free energy. The factor $\Np$ accounts for the fact that the number of sites at which a nucleus can form is linear in polymer length. The attempt rate depends on the rate at which adsorbed and/or free subunits associate with a polymer-bound partial capsid intermediate, which depends in part on the rates of correlated polymer-subunit motions and subunit diffusion along the polymer. If nucleation is dominated by association of subunits that are already adsorbed onto the polymer, the rate can be expressed as $\tnuc^{-1} \approx f^\prime c_1 c_{n-1}$ with $f^\prime$ a rate constant for polymer-adsorbed subunits. This scaling was found to be consistent with the simulation data in Ref.~\cite{Kivenson2010}. In performing this analysis, it is important to note that the critical nucleus size $n$ can depend on interaction free energies and subunit concentrations (see Refs.~\cite{Zandi2006,Hagan2010}). We have not performed a statistical analysis using committor probabilities \cite{Kivenson2010,Dellago2002} but for the conditions studied in this manuscript, critical nucleus sizes for assembly on the polymer appear to fall in the range $3 \lesssim \nnuc \lesssim 5$ (see \ref{sec:freeenergy}).

\section{Estimates of Binding Free Energies}
\label{sec:freeenergy}

{\bf Capsid subunit-subunit binding free energies.} In order to estimate the free energy subunit-subunit binding, we performed simulations of subunits with attractors on only one of the three edges, so that only dimerization was possible. We measured the dimer-monomer dissociation constant $\Kd$ in the absence of polymer for a range of subunit concentrations and binding energies $\ecc$. The free energy for binding along a single interface (which involves up to six attractors on each subunit) is then given by $\gcc=-\kt \ln (\Kd^{-1} \css)$ where $\css=8 \sigma\rsub{b}^{-3}$ is the standard state concentration that maps to the conventional choice of 1 M (see  section~\ref{sec:lengthScales}). The resulting free energy can be expressed as $\gcc \approx - 3.5 \ecc - T\scc$, with the binding entropy $\scc=-12.4\kb$. The binding entropy arises from rotational entropy loss and the fact that the subunit-subunit attraction range $\sigma\rsub{a}$ is smaller than the standard state length scale $\sigma\rsub{b}/2$. We calculated the binding entropy analytically for a similar model in Ref.~\cite{Hagan2006}; for further discussion also see Refs.\cite{Erickson1981,Ben-tal2000}.

We can obtain an upper bound on the free energy of larger capsid structures by noting that the binding entropy for dimerization $\scc$ is a lower bound for the entropy lost by a subunit with multiple bonds. Furthermore, the majority of the entropy is lost upon making the first bond, because the contacts are so stereospecific.  A rough estimate for the free energy per subunit of a well-formed model capsid with $\Nc=20$ subunits is therefore $G_\mathrm{capsid} \gtrsim 3/2 \Nc (3.5 \ecc) + (\Nc-1) T \scc$ to give the free energy per subunit $g_\mathrm{capsid}=-9.2\kbt$ at $\ecc=4.0\kbt$. We note that, despite the fact that forming a capsid is thermodynamically favorable at these parameters, capsids do not spontaneously assemble in our simulations until $\ecc=5.0\kt$ because of a large nucleation barrier -- the smallest (weakly) favorable structure is a pentamer. For $\ecc=5.0\kt$ our estimate gives $g_\mathrm{capsid}=-14.5 \kt$, which is consistent with experimental values for the free energy per subunit at which capsids spontaneously assemble \cite{Ceres2002,Parent2006,Zlotnick2003a}. 

\begin{figure}
  \includegraphics[width=0.5\textwidth]{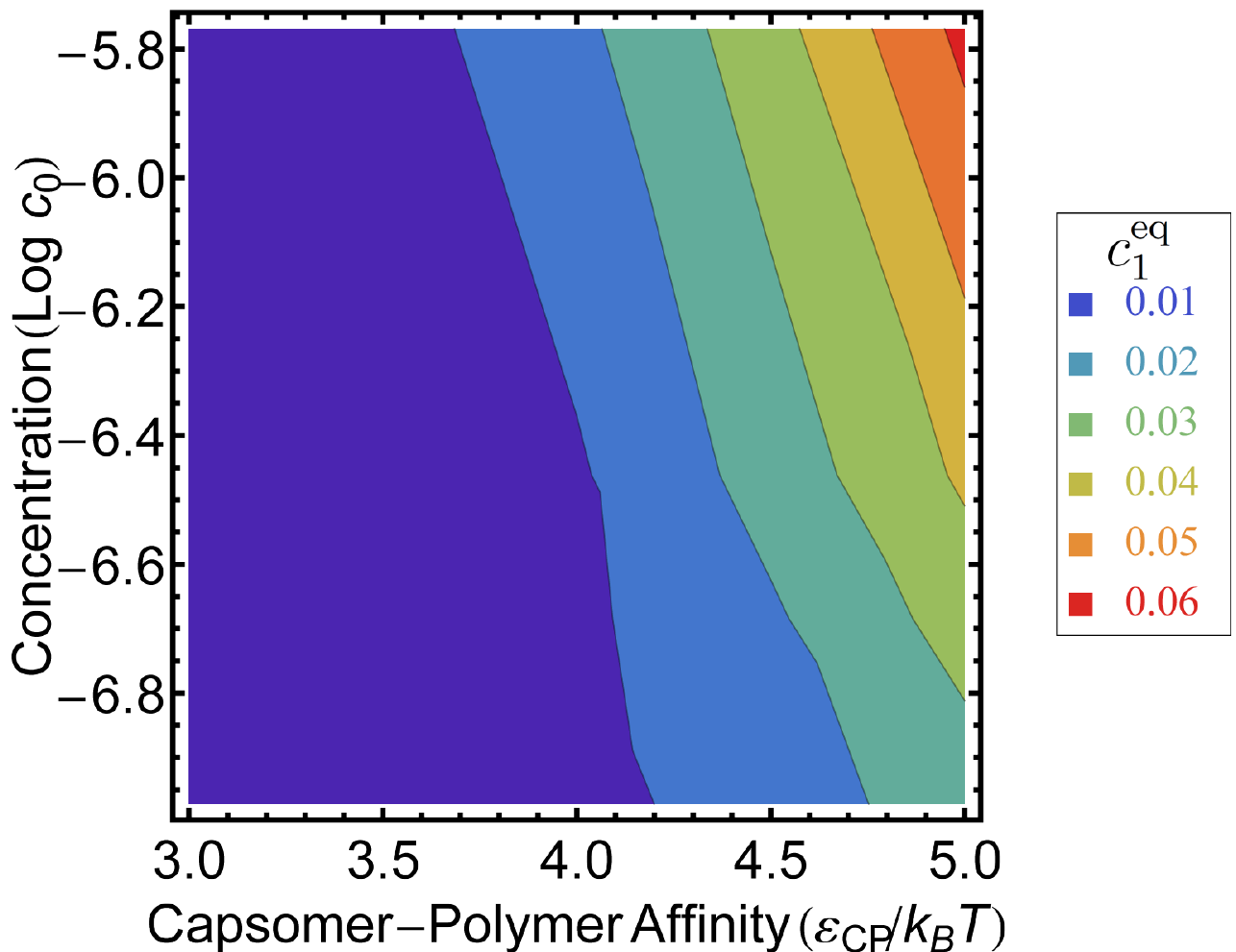}
  \caption{The driving force for subunits to adsorb on to the polymer is revealed by $\csurf$, the equilibrium one-dimensional concentration of subunits on the polymer in the absence of capsomer-capsomer attractions ($\ecc=0$). $\csurf$ is measured as the average number of adsorbed subunits divided by the polymer length, and shown as functions of $\ecp$ and $\logc$. }
  \label{fig:csurf}
\end{figure}

{\bf Capsid-polymer binding free energy.} We estimate the polymer-capsomer binding free energy by performing simulations in which the capsomer-capsomer binding energy is set to zero $\ecc = 0.0$ (figure \ref{fig:csurf}). We can then extract the binding free energy from the formulation given by McGhee and von Hippel for the binding of a ligand to a uniform polymer when each ligand occupies more than one binding site\cite{McGhee1974}. We find that the free energy of binding for intermediates binding energies $3.0\kbt \leq \ecp \leq 5.0\kbt$ is given by $\gcp \sim -1.9 \ecp -T\scp$ with $\scp=-7.4 \kb$. Again note that binding of a single subunit to the polymer, with $\Kd=4$ mM, is unfavorable at the default concentration we consider, $c_0=6.25 \times 10^{-4} \sigma\rsub{b}^{-3}$ or $80 \mu$M. 

\section{Further information}
\label{sec:further}

\begin{figure}
\subfloat[] {
  \includegraphics[width=0.45\textwidth]{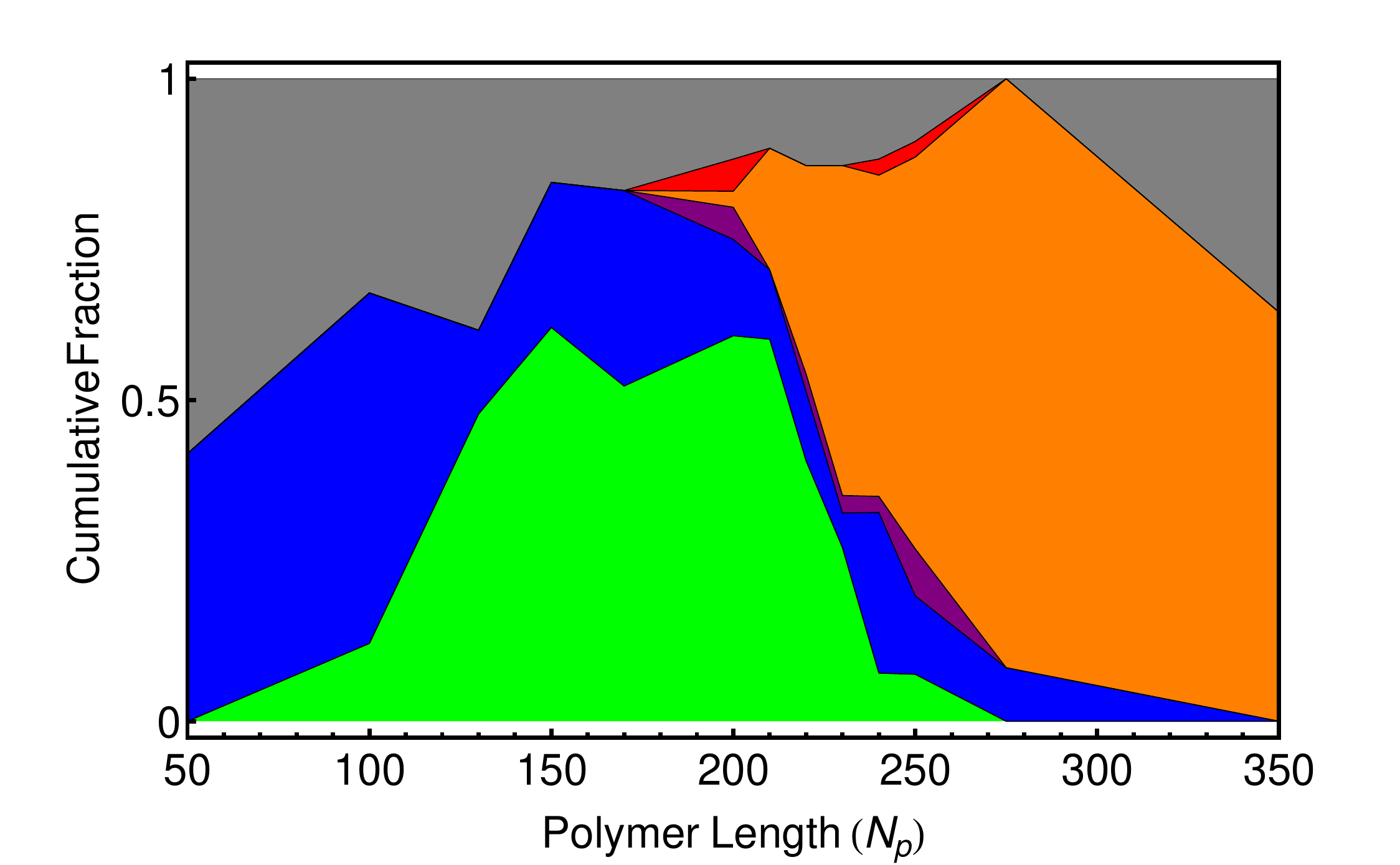}
  \label{fig:stack3.5}
}
\subfloat[] {
  \includegraphics[width=0.45\textwidth]{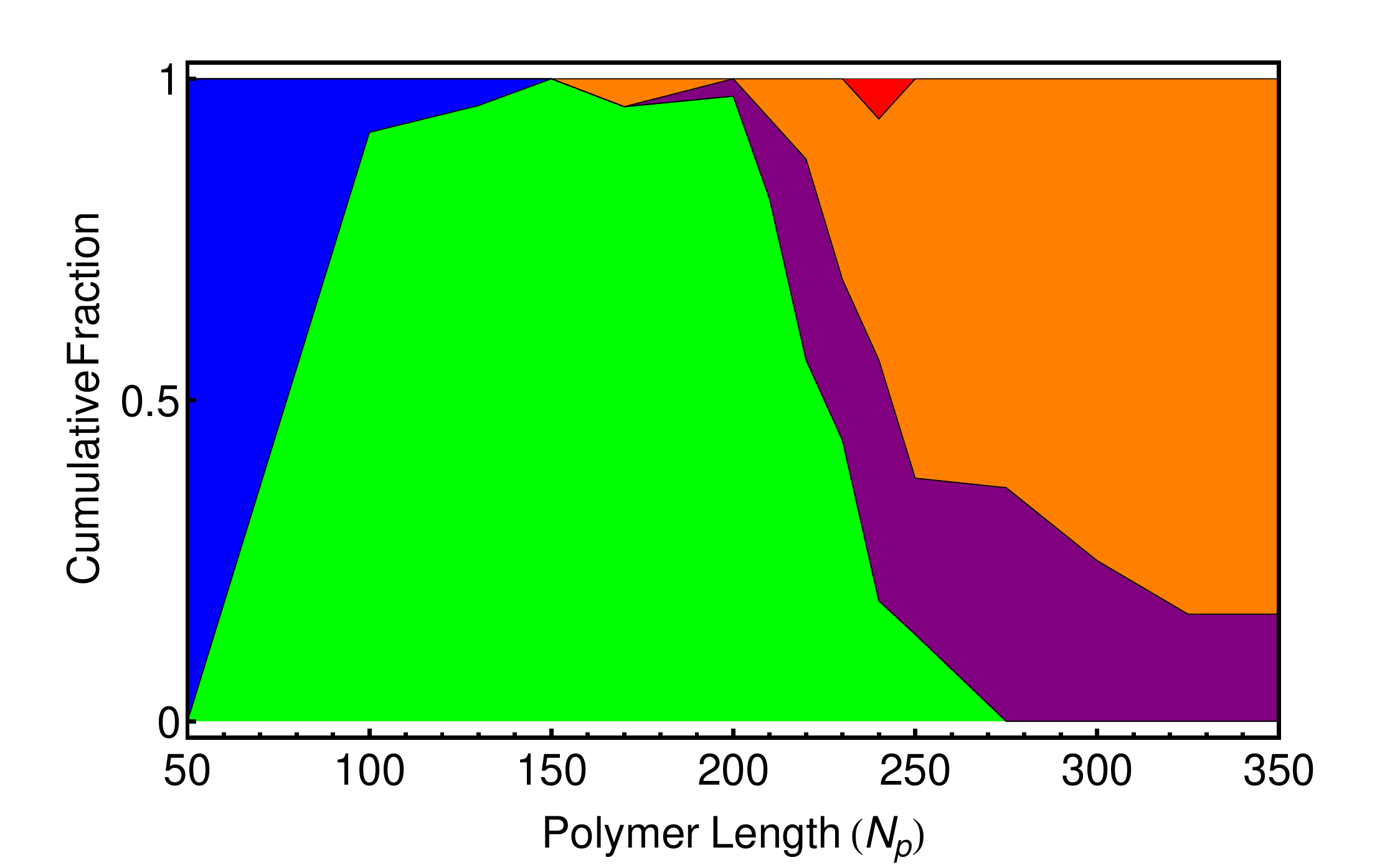}
  \label{fig:stack4.0}
}
\\
\subfloat[] {
  \includegraphics[width=0.45\textwidth]{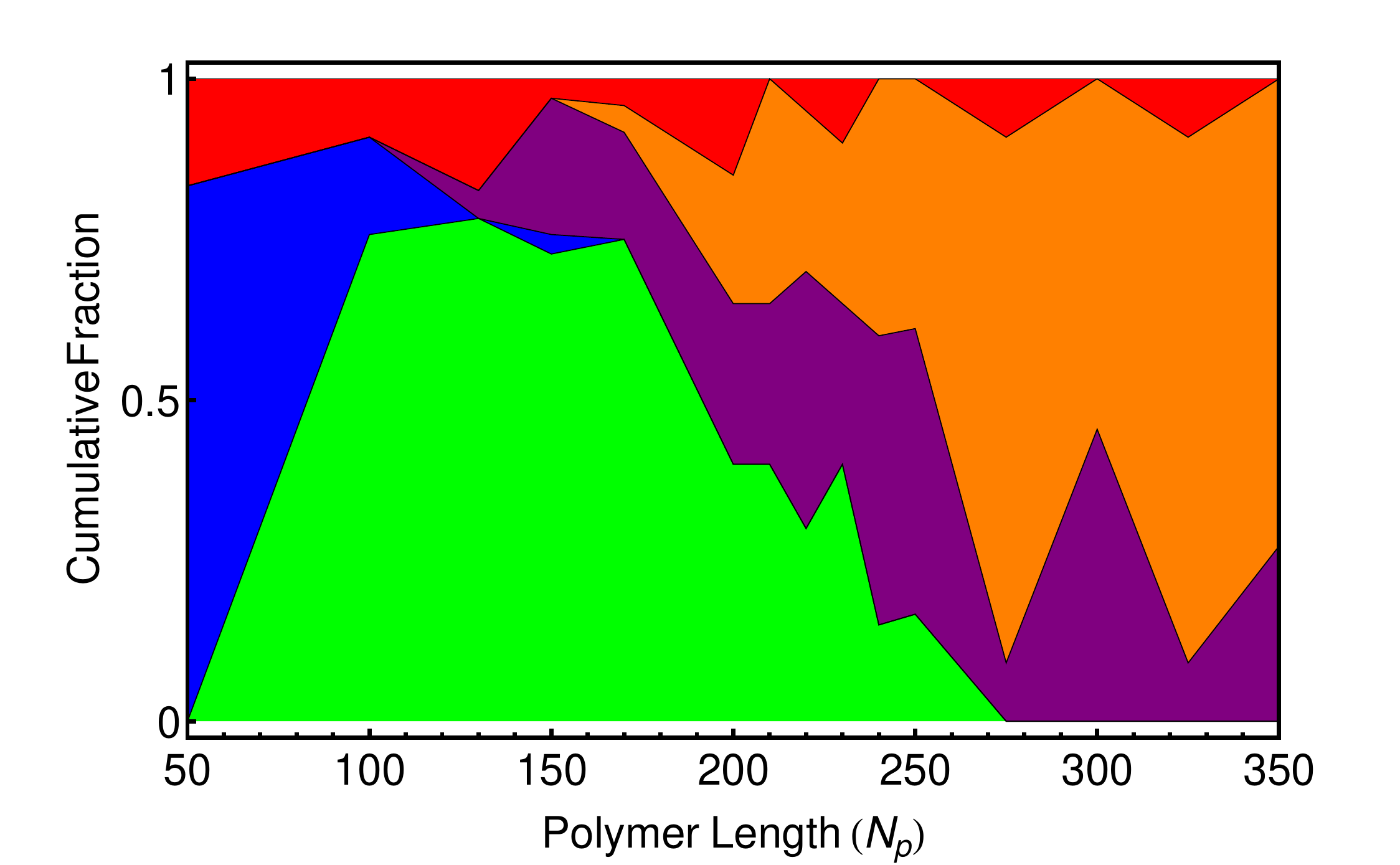}
  \label{fig:stack4.5}
}
\subfloat[] {
  \includegraphics[width=0.45\textwidth]{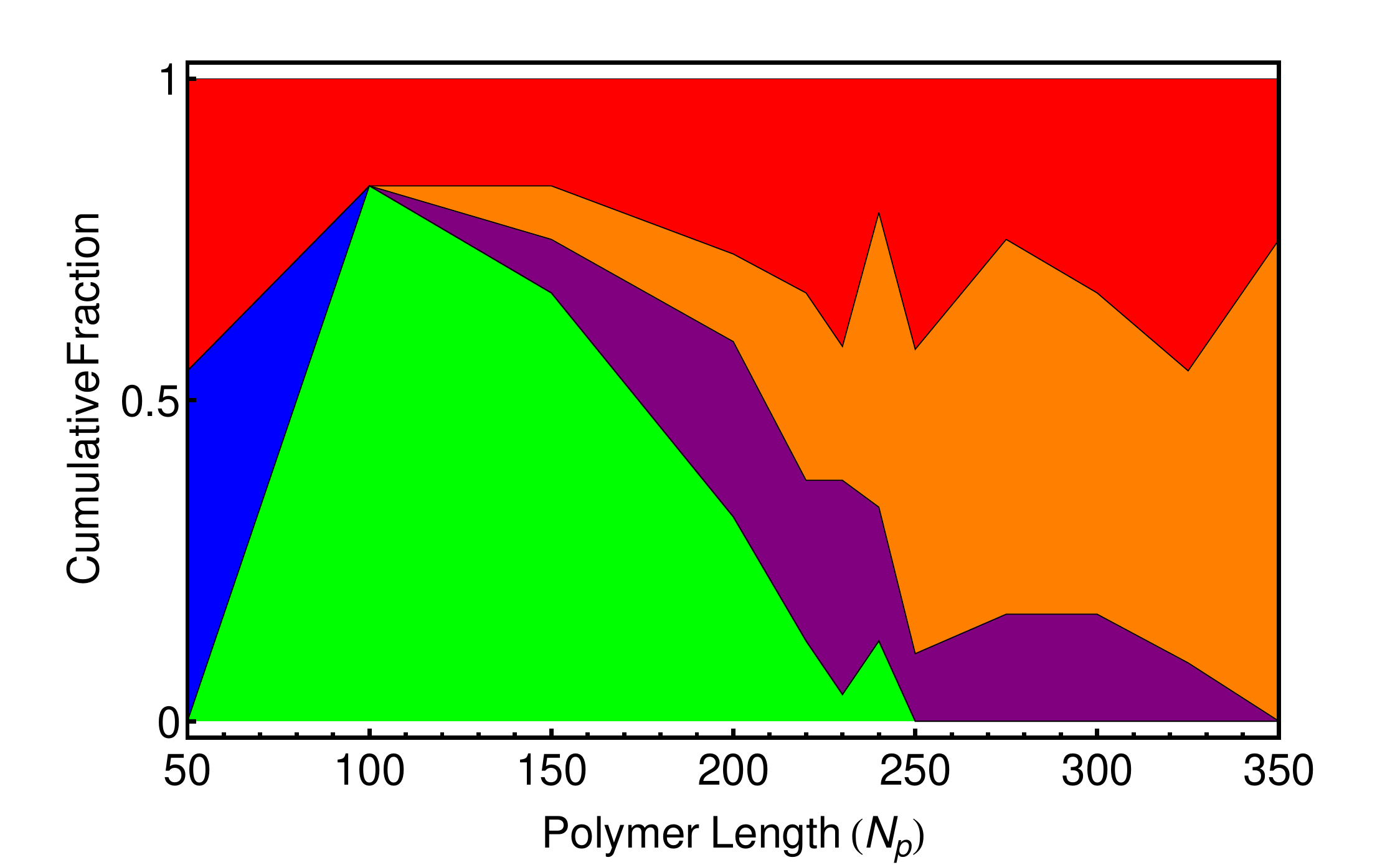}
  \label{fig:stack5.0}
}
\caption{The fraction of trajectories that end in each outcome are shown in cumulative plots as a function of $\Lp$ for $\ecp = \{3.5,4.0,4.5,5.0\}$, for (a)-(d) respectively.  The height of each color corresponds to the fraction of trajectories resulting in that outcome, color-coded according to the legend in figure \ref{fig:kpdleg}. The spike in at $\Lp\sim300$ in {\bf (c)} corresponds to a large yield of size 30 defective capsids, examples of which are pictured in the bottom row of figure \ref{fig:hexamers}.
}
\label{fig:cumulative}
\end{figure}

\begin{figure}
\includegraphics[width=0.9 \textwidth]{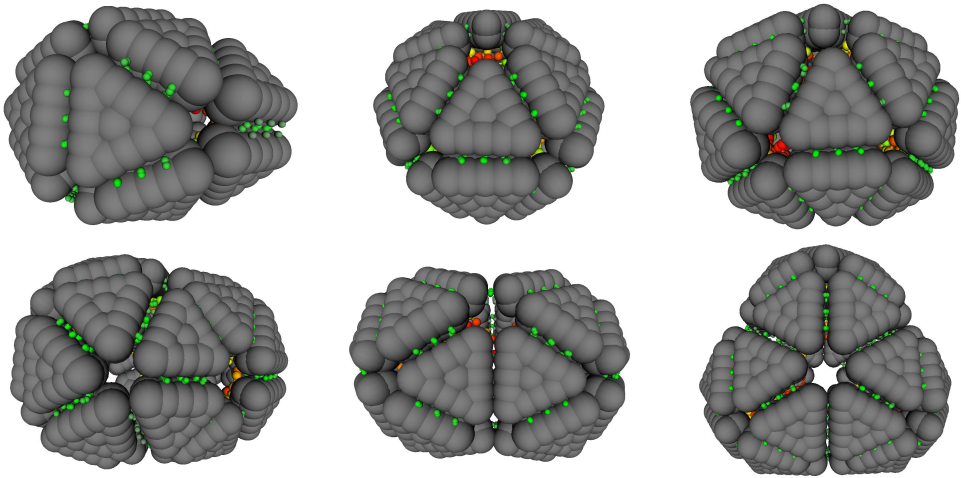}
\caption{ Examples of common malformed but closed capsids. The top row shows the single dominant morphology for sizes 22, 24 and 26. For sizes 24 and 26, the dislocations (2 in the former case, 3 in the latter) relieve strain by arranging themselves at opposite poles of the 2 and 3 fold symmetry axes, respectively. In the bottom row are the 3 most prevalent morphologies for malformed capsids of size 30, for which more strain-relieving arrangements of hexamers are possible.}
\label{fig:hexamers}
\end{figure}

\begin{figure}
\subfloat[] {
    \includegraphics[width=0.3 \textwidth]{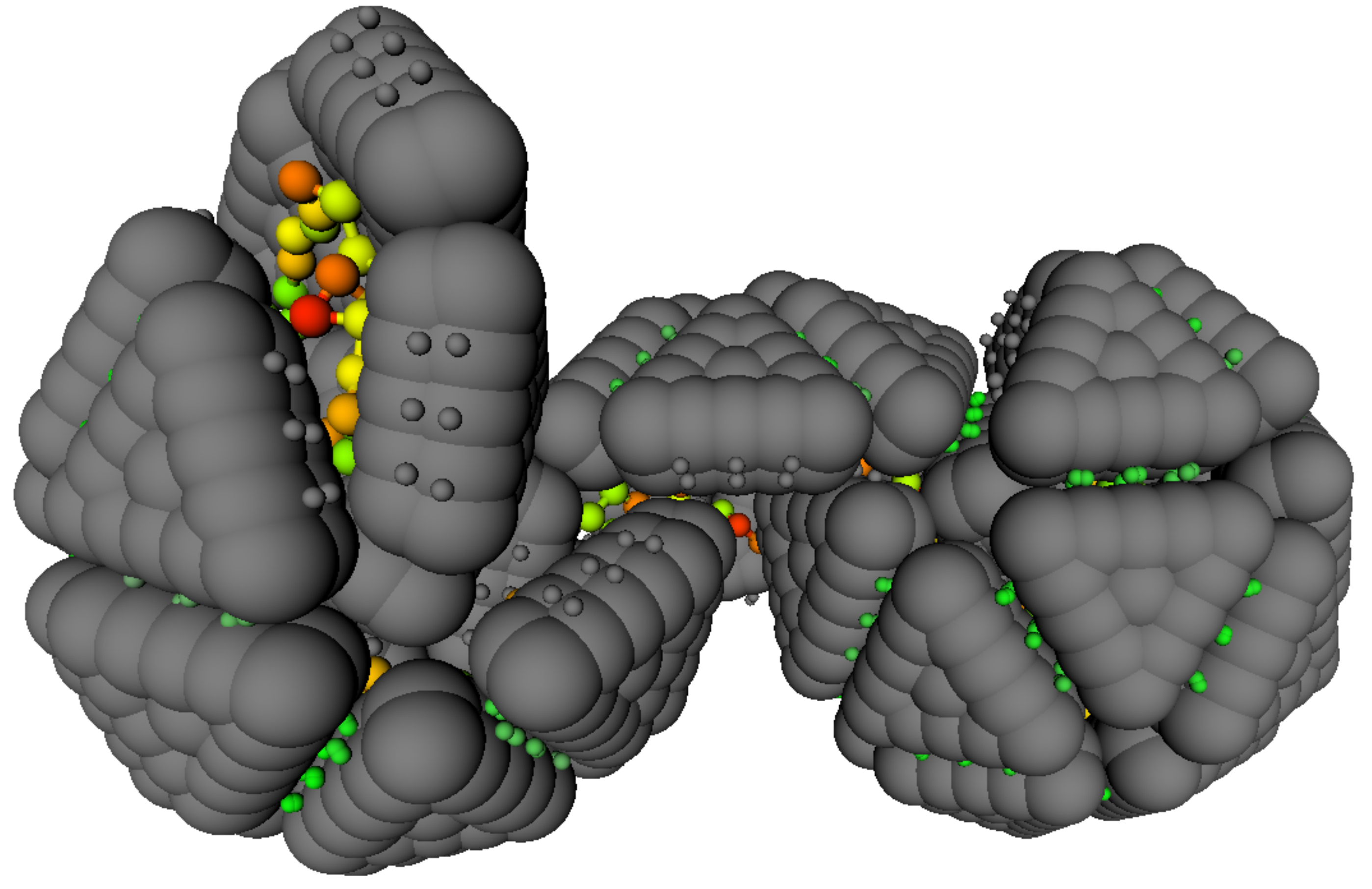}
}
\subfloat[] {
    \includegraphics[width=0.3 \textwidth]{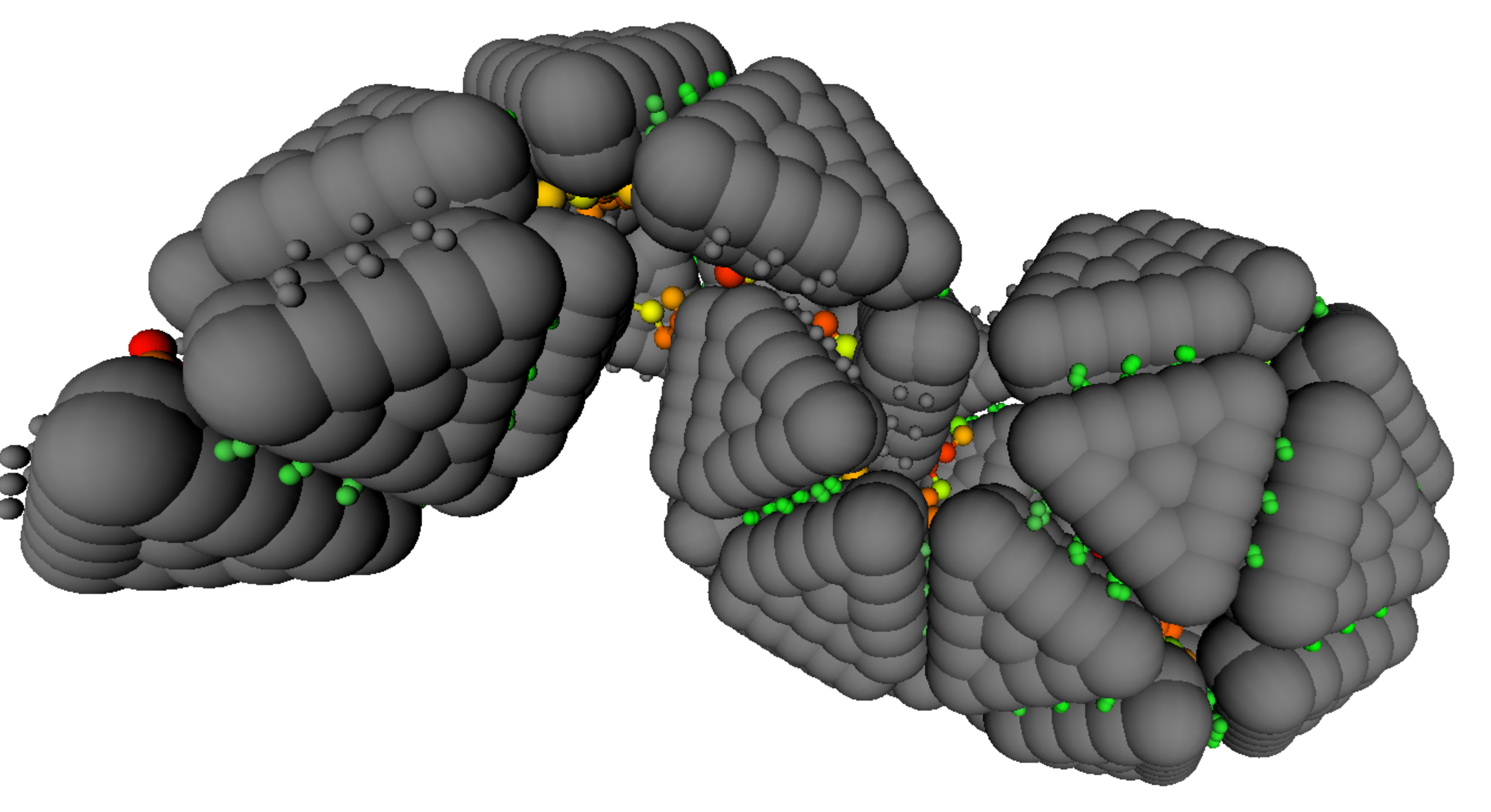}
}
\subfloat[] {
    \includegraphics[width=0.3 \textwidth]{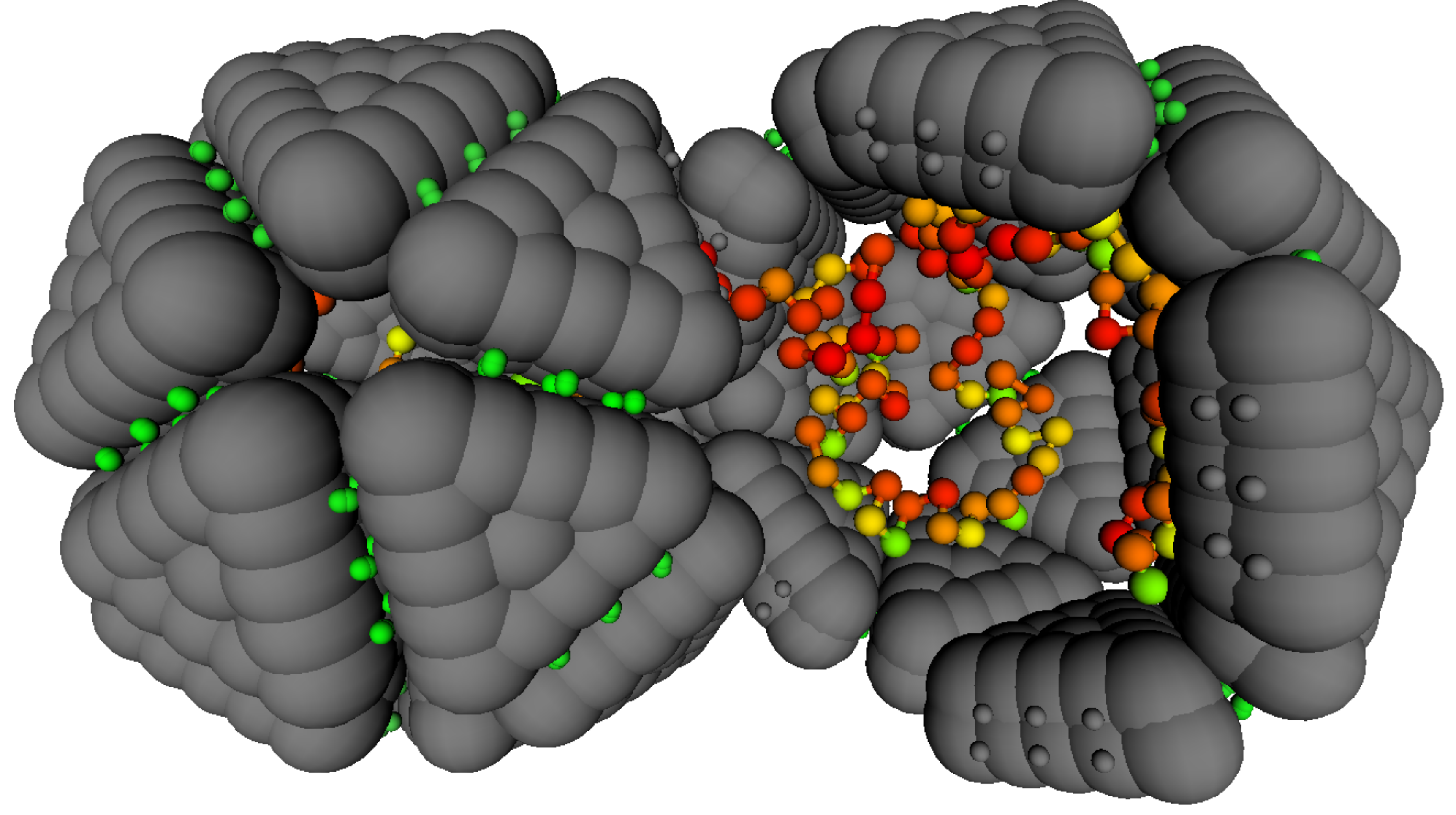}
}
\caption{ Typical disordered assembly products for high capsomer-polymer affinities $\ecp \geq 5.0$.}
\label{fig:disordered}
\end{figure}
\begin{figure}
\subfloat[] {
    \includegraphics[width=0.45 \textwidth]{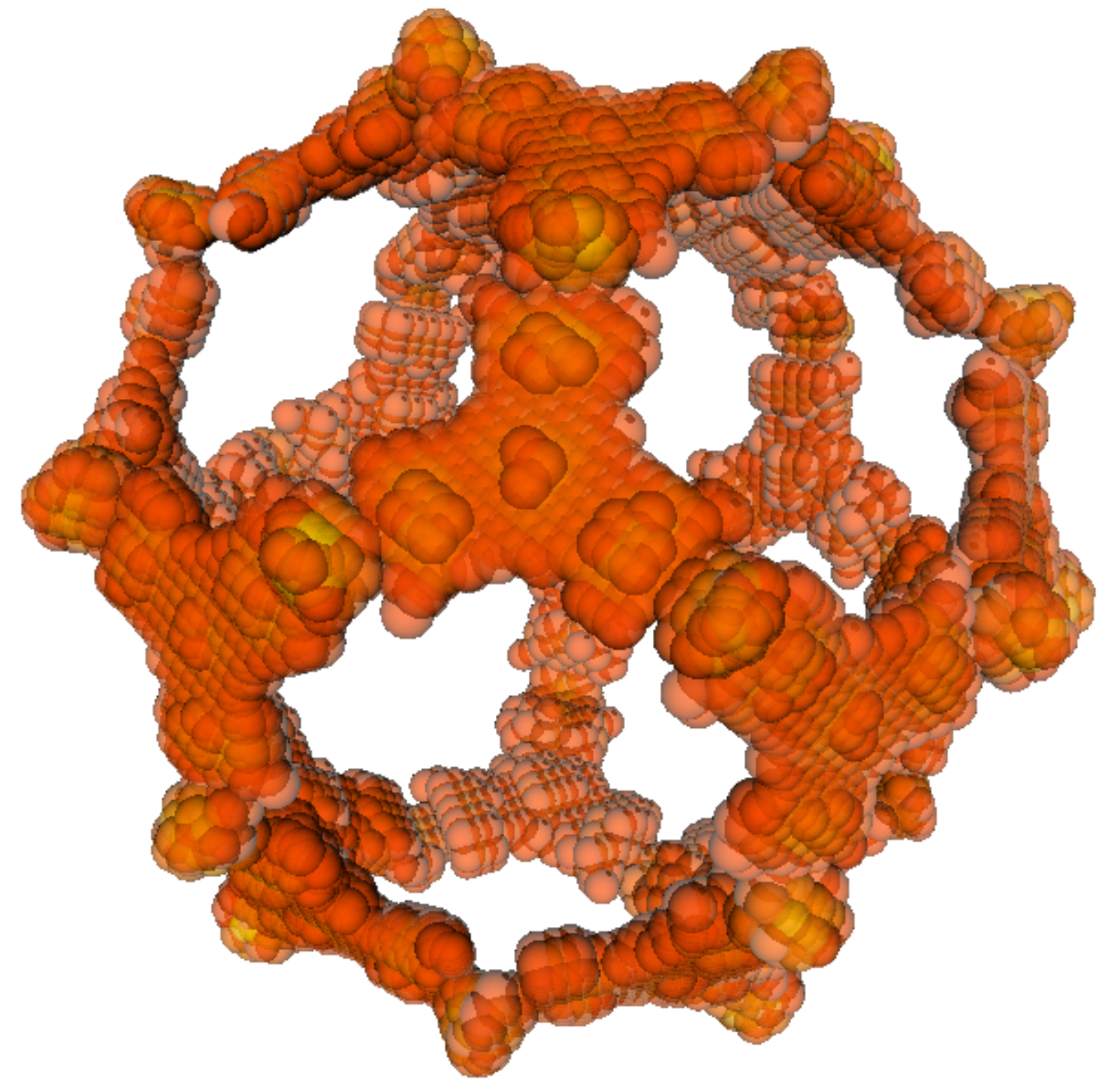}
}
\subfloat[] {
    \includegraphics[width=0.45 \textwidth]{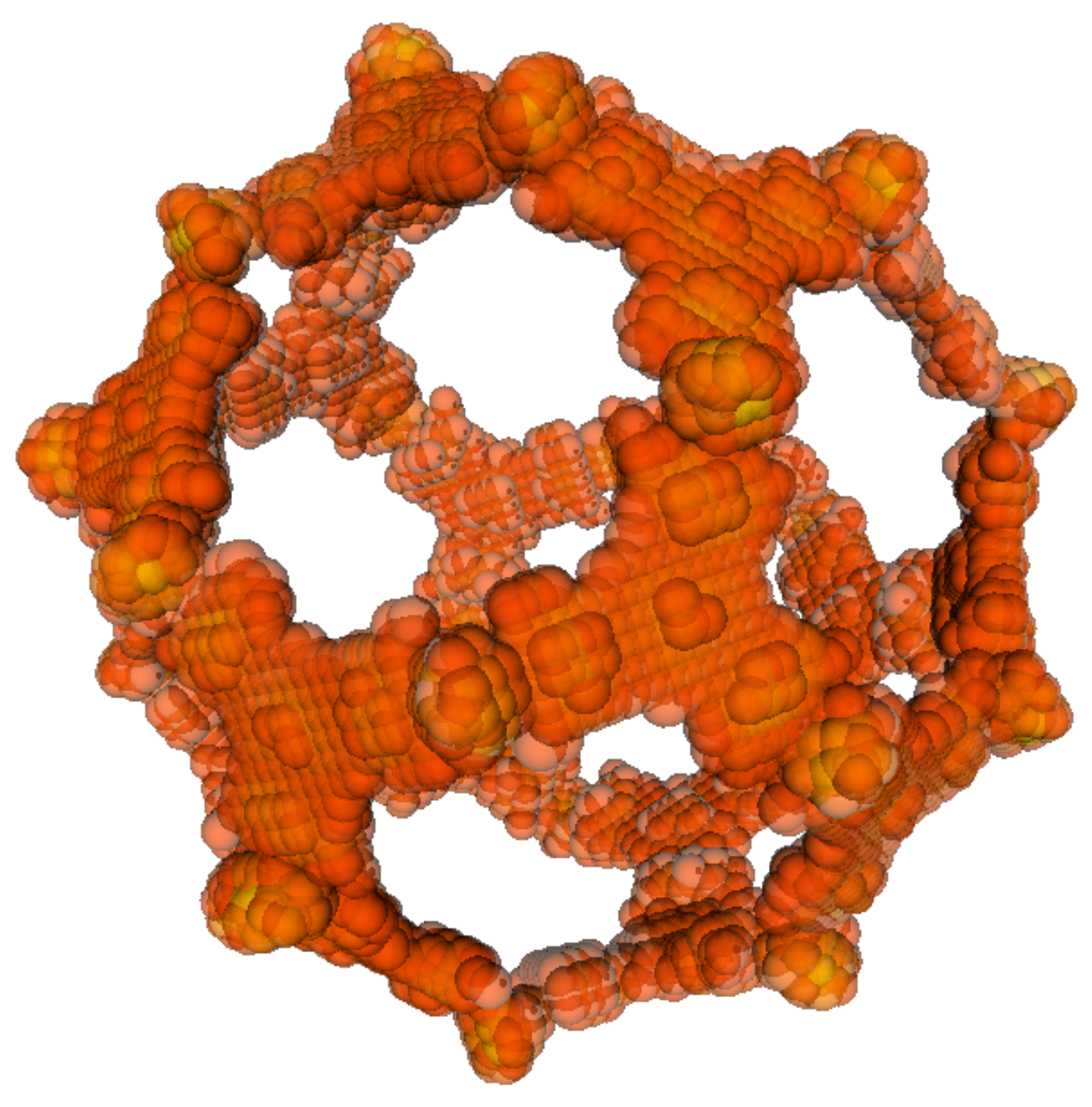}
}
\caption{Visualization of the polymer density. The polymer density is averaged over a large number of successful assembly trajectories after completion, for a polymer with length $\Lp=150$. Densities are averaged over the threefold symmetry of the capsomer, but not over the 20-fold symmetry group of the completed capsid.}
\label{fig:icospoly}
\end{figure}

\end{document}